\documentclass[aps,prd,onecolumn,nofootinbib,amsmath,amssymb]{revtex4}

\usepackage[utf8]{inputenc}
\usepackage{epsfig}
\usepackage{graphicx}
\usepackage{subfig}
\usepackage{hyperref}
\usepackage{amsfonts}
\usepackage{ulem}
\usepackage[titletoc,toc,title]{appendix}

\def\Mp{M_{n,{\bf p}}}

\def\Tr{{\rm Tr\,}}
\def\sumint{\sum\!\!\!\!\!\!\!\int\,}

\begin{document}

%\date{}
\title{Thermodynamics of an exactly solvable confining quark model}

\author{M.~S.~Guimaraes\footnote{msguimaraes@uerj.br}, B.~W.~Mintz\footnote{bruno.mintz.uerj@gmail.com} and L.~F.~Palhares\footnote{leticiapalhares@gmail.com} }

\affiliation{
Departamento de F\'{\i}sica Te\'orica, Universidade do Estado 
do Rio de Janeiro, 20550-013, Rio de Janeiro, RJ, Brazil\\
}

\begin{abstract}
 The grand partition function of a model of confined quarks is exactly calculated at arbitrary temperatures and quark chemical potentials. 
 The model   is inspired by a softly BRST-broken version of QCD and possesses a quark mass function compatible with nonperturbative 
 analyses of lattice simulations and Dyson-Schwinger equations.   Even though the model is defined at tree level, we show that it 
 produces a nontrivial and stable thermodynamic behaviour at any temperature or chemical potential. Results for the pressure, 
 the entropy and the trace anomaly as a function of the temperature are qualitatively compatible with the effect of nonperturbative 
 interactions as observed in lattice simulations. The finite density thermodynamics 
 is also shown to contain nontrivial features, being far away from an ideal gas picture.
 \end{abstract}

\maketitle

%%%%%%%%%%%
\section{Introduction}

The study of the infrared behavior of non-Abelian gauge theories, like Quantum Chromodynamics, presents a remarkable task for 
theoretical physics. The strongly-coupled nonperturbative nature of infrared phenomena greatly undermines our ability to access 
the physical properties of the theory in an analytical way. One general question raised in this context concerns the description of the degrees 
of freedom in a confining theory. Confinement of the perturbative excitations associated with the fundamental fields means, among other features,
 that these degrees of 
freedom do not appear in the physical spectrum of the theory as asymptotic states. The physical spectrum in this case is given by bound states 
associated with composite fields of the theory. The fundamental problem is thus to understand how the perturbative degrees of freedom disappear 
from the spectrum   and how this is reflected in an action formulation written in terms of them. 

\vspace{.1in}

A very important tool in identifying the physical space of states in a gauge theory is the BRST symmetry \cite{Becchi:1975nq,Tyutin:1975qk}. 
In a BRST invariant theory, the physical states are defined by the cohomology of the nilpotent BRST operator that acts on the fundamental 
fields.   However, this is strictly valid only at the perturbative level. In fact, the whole construction of the BRST formalism in a gauge theory 
can be traced back to the Faddeev-Popov procedure of gauge fixing. Since it has long been known that this procedure ceases to be valid at the 
nonperturbative level \cite{Gribov:1977wm}, there is no strict necessity 
to expect that the BRST symmetry survives in the infrared regime. 
Indeed, there is strong evidence coming from lattice studies that the BRST symmetry is broken at low energies \cite{Cucchieri:2014via}. 
This is in tune with the expectations for a confining theory: the survival of the original BRST symmetry would allow for a definition 
of the physical spectrum, by its cohomology, in terms of the fundamental fields, thus contradicting confinement. 

\vspace{.1in}

It is  therefore expected that the description of the original perturbative degrees of freedom is modified in the infrared. 
 This seems to be exactly what happens in pure gauge theories according to lattice results concerning the Yang-Mills gauge field 
and ghost propagators in the Landau gauge \cite{Cucchieri:2007rg,Cucchieri:2008fc,Cucchieri:2011ig}:  
positivity violation is observed for the gluon two-point function. It is well-known \cite{Peskin:1995ev,Weinberg:1995mt} that the two-point function 
of a physical field is directly related to the probability of propagation of the associated asymptotic state. Therefore, positivity violation, as observed
for lattice gluons, hinders the probabilistic interpretation of the two-point function in terms of propagating particles, being consistent with confinement
in the sense of absence of colored states from the spectrum. This is the confinement criterion we adopt (more discussion can be found in e.g.
the reviews \cite{Alkofer:2000wg,Brambilla:2014jmp}).

The  numerical data 
also point to gluon propagator that tends to a constant value in the infrared,  while the 
ghost propagator has the behavior of a free particle. In the context of the Schwinger-Dyson equations, this is called the 
decoupling solution \cite{Aguilar:2008xm}. The theoretical description of these results in the continuum 
has been proposed to be provided by an effective action in the so-called refined Gribov-Zwanziger (RGZ) formalism 
\cite{Dudal:2007cw, Dudal:2008sp, Dudal:2011gd}. The RGZ action is renormalizable and displays a softly broken BRST symmetry. 
One very important feature of this formalism is the guarantee that the quantum ultraviolet behavior of the theory is 
under control   and the -- very successful -- original perturbative Yang-Mills framework is recovered at high energies.

\vspace{.1in}

The RGZ action describes a pure gluonic theory and its construction has the geometrical guidance of the original Gribov observations 
about the configuration space of the gauge fields. Its action effectively describes the restriction of the path integral measure to 
a region without the infinitesimal Gribov gauge copies, while also taking into account the formation of condensates. The resulting action has tree-level 
propagators that describe very well the lattice data at zero and finite temperature \cite{Cucchieri:2010xr,Bornyakov:2011fn} and can be used 
to obtain consistent estimates of the masses of glueball states with different quantum numbers  \cite{Dudal:2010cd, Dudal:2013wja}. 

\vspace{.1in}

An interesting development is that lattice data on propagators suggest a similar behavior for the  gauge-interacting 
matter fields, displaying positivity violation \cite{Parappilly:2005ei}, as observed for the gluon fields. Furthermore, the theoretical description of 
these degrees of freedom in the infrared has been conjectured to have the same structure as for the gluon, with an action whose 
tree-level propagators reproduce very well the lattice data,  also displaying softly broken BRST symmetry \cite{Capri:2014bsa}.  It should 
be noticed, however, that in the case of matter fields a geometrical interpretation of the modified action as a restriction in the 
configuration space of the fields is lacking.  

\vspace{.1in}

This set of affairs led us to expect that a  possible signature of confinement would be the soft breaking of the BRST symmetry. In fact this 
relation has been explored in many works \cite{Baulieu:2008fy, Dudal:2009xh, Sorella:2009vt, Sorella:2010it, Capri:2010hb, Dudal:2012sb, Reshetnyak:2013bga}. The idea is that confined degrees of freedom are described by an effective action with a  BRST-breaking soft term.  Interestingly, the breaking of the BRST invariance may have a connection with chiral symmetry breaking, since the currently known formulations of soft BRST breaking in the quark sector always imply chiral symmetry breaking \cite{Baulieu:2009xr, Dudal:2013vha} and thus dynamical mass generation in the deep infrared seems to be intimately related to BRST breaking  \cite{Capri:2014bsa}. In this sense, one may hope that the investigation of the nonperturbative breaking 
of the BRST symmetry may shed some light on a long-standing puzzle in the QCD phase structure: the apparent and unexplained link 
between confinement and chiral symmetry breaking.

\vspace{.1in}

Several effective models of QCD present in the literature address the relation between chiral symmetry breaking and confinement. 
Two important prototype models extensively explored to address the question of chiral symmetry breaking in the strong interactions 
are the Nambu-Jona-Lasinio (NJL) and the Quark Meson (QM) 
models \cite{Nambu:1961tp,GellMann:1960np}, generically called chiral models. On one hand, chiral symmetry breaking and restoration is, 
in chiral models, a result of the interaction of the quark sector either to itself (in a four-fermion vertex, as in the NJL model) or to 
meson fields (as in the QM model). On the other hand, as an attempt to take into account the physics of confinement, one can introduce a 
Polyakov loop field coupled to quarks \cite{Meisinger:1995ih,Fukushima:2003fw,Megias:2004hj,Ratti:2005jh,Schaefer:2007pw,Schaefer:2009ui}. 
Within the Polyakov loop extended NJL (PNJL) and QM (PQM) models, the Polyakov loop field effectively encodes the gauge field as an external background 
that interacts with the 
quark degrees of freedom, so that the deconfinement transition may be studied in this context as a (static) Landau-Ginzburg system coupled to the chiral model. 
In spite of their interaction with the Polyakov loop, the quark degrees of freedom in both the PNJL and PQM models physically correspond to asymptotic 
states, not to confined ones. This is so because their two-point functions only have poles on the real axis and thus respect the Osterwalder-Schrader's axiom of 
reflection positivity \cite{Osterwalder:1973dx}, guaranteeing the establishment of a standard probabilistic description of propagation of asymptotic quarks, in contradiction with real-world QCD.

\vspace{.1in}

A possible way to extend the NJL model is to include a nonlocal current-current interaction kernel 
\cite{General:2000zx,Contrera:2010kz,Benic:2012ec,Benic:2013eqa}. As a result, complex quark 
masses may appear, indicating confinement in the sense we discussed before. However, due to its four-fermion interaction, the nonlocal 
NJL model is not renormalizable, similarly to its local version. Besides, it also leads to unstable thermodynamic behavior 
\cite{Benic:2012ec,Benic:2013eqa,Loewe:2014vna}. 

\vspace{.1in}

  In the last years, some works have explored the thermodynamic properties of softly BRST broken pure gauge systems (see, for example,  
\cite{Zwanziger:2004np,Zwanziger:2006sc,Lichtenegger:2008mh,Fukushima:2013xsa, Reinhardt:2012qe,Reinhardt:2013iia}).    
In this paper, we explore thermodynamic properties 
of a model of quarks with soft BRST breaking. As discussed above, this model is expected to describe the infrared properties of confined  (positivity-violating) quarks, 
while keeping compatibility with ultraviolet QCD properties. Indeed, the analytical propagator of the model fits well the available lattice data \cite{Dudal:2013vha} and the model has been proven to be renormalizable \cite{Baulieu:2009xr}, reducing to perturbative quarks in the ultraviolet regime.  Our goal here is to show not only that soft BRST breaking in the quark sector  implies 
a well-defined   macroscopic behaviour, but also that the tree level model is capable of predicting nontrivial features, being in general 
qualitatively compatible with the effect of nonperturbative interactions as observed in lattice data. Furthermore, the instabilities present in the 
nonlocal NJL model are absent from our setup.

\vspace{.1in}

This work is organized as follows.   Section II presents the action defining the model. In Section III the exact partition function is 
computed within the imaginary-time formalism. In section IV  we obtain the thermodynamic quantities and present our results for the distribution 
function, the pressure, the entropy, and the trace anomaly. Our conclusions are discussed in Section V.

%%%%%%%%%%%
\section{The action of the quark model with soft BRST breaking}

Following the discussion of \cite{Baulieu:2009xr}, let us briefly review how a model with soft breaking of the BRST symmetry can be 
obtained from an extension of the QCD lagrangian. We start from the gauge-fixed lagrangian density in an euclidean space of dimension 4,
\begin{equation}\label{eq:QCD-lagrangian}
   {\cal L}_{QCD} = \frac14 F_{\mu\nu}^aF_{\mu\nu}^a + \bar\psi_\alpha^i[i(\gamma_\mu)_{\alpha\beta}D_\mu^{ij} - m_0\delta_{\alpha\beta}\delta^{ij}]\psi_\beta^j 
                + ib^a\partial_\mu A_\mu^a + \bar c^a \partial_\mu D_\mu^{ab}c^b, 
\end{equation}
where $F_{\mu\nu}^a = \partial_\mu A_\nu^a - \partial_\nu A_\mu^a + g f^{abc}A_\mu^b A_\nu^c$ is the field 
strength and $D^{ij}_\mu = \delta^{ij}\partial_\mu - ig(T^a)^{ij}A_\mu^{a}$ is the covariant derivative. The 
indices $(a,b,\dots,h) = 1,\dots,N_c^2-1$ correspond to adjoint representation and $(i,j, \dots ) = 1, \dots, N_c$ 
correspond to fields in the fundamental representation of $SU(N_c)$. The greek indices $(\alpha,\beta,\dots) = 1,2,3,4$
denote spinor indices while $(\mu,\nu,\dots) = 1,2,3,4$ regard euclidean space indices. For more details on 
notational conventions, the reader is referred to Ref. \cite{LeBellac:book}.

One very important feature of (\ref{eq:QCD-lagrangian}) is its invariance under BRST transformations \cite{Becchi:1975nq,Tyutin:1975qk},

 \begin{eqnarray}
  && sA_\mu^a=-D_{\mu}^{ab}c^b \nonumber\\
  && s\psi_\alpha^i = -ig c^a (T^a)^{ij}\psi^j_\alpha \nonumber\\
  && s\bar\psi_\alpha^i = -ig\bar\psi_\alpha^j c^a(T^a)^{ji} \nonumber\\
  && sc^a = \frac12 g f^{abc} c^b c^c \nonumber\\
  && s\bar c^a = ib^a \nonumber\\
  && s b^a = 0
 \end{eqnarray}
which is crucial for its multiplicative renormalizability. As an intermediate step in the definition of the model, 
let us introduce two BRST doublets $(\xi^i,\theta^i)$ and $(\eta^i,\lambda^i)$ of auxiliary fields (and their hermitian 
conjugate fields) that transform as
\begin{eqnarray}
 && s\xi_\alpha^i = \theta_\alpha^i,\;\;\;\;s\theta_\alpha^i = 0\nonumber\\
 && s\eta_\alpha^i = \lambda_\alpha^i,\;\;\;s\lambda_\alpha^i = 0.
\end{eqnarray}
 The addition of the BRST invariant action
\begin{eqnarray}\label{eq:auxiliary-fields}
    S_{\xi\lambda} &=& s\int d^4x\left[ -\bar\eta^i_\alpha\partial^2\xi_\alpha^i + \bar\xi^i_\alpha\partial^2\eta_\alpha^i + m^2(\bar\eta^i_\alpha\xi_\alpha^i - \bar\xi_\alpha^i\eta_\alpha^i) \right]\nonumber\\
    &=& \int d^4x\left[-\bar\lambda_\alpha^i\partial^2\xi_\alpha^i - \bar\xi^i_\alpha\partial^2\lambda_\alpha^i - \bar\eta_\alpha^i\partial^2\theta_\alpha^i + \bar\theta_\alpha^i\partial^2\eta_\alpha^i + m^2\left(\bar\lambda^i_\alpha\xi_\alpha^i + \bar\xi_\alpha^i\lambda_\alpha^i + \bar\eta_\alpha^i\theta_\alpha^i - \bar\theta_\alpha^i\eta_\alpha^i\right) \right]
\end{eqnarray}
does not change the physical content of the theory. This can be easily checked by integrating the auxiliary fields, which 
trivially gives $1$. However, if one also adds to the action the coupling 
\begin{equation}\label{eq:BRST-breaking}
    S_M = \int d^4x\left[M_1^2(\bar\xi^i_\alpha\psi_\alpha^i + \bar\psi_\alpha^i\xi_\alpha^i) - M_2(\bar\lambda_\alpha^i\psi_\alpha^i + \bar\psi_\alpha^i\lambda^i_\alpha)\right]
\end{equation}
between the auxiliary fields and the matter fields, the resulting action $S_{tot}=S_{QCD}+S_{\xi\lambda}+S_M$ 
is no longer BRST invariant. However, as shown in \cite{Baulieu:2009xr}, this does not destroy the renormalizability 
of the theory once it corresponds to a {\it soft} breaking of the BRST symmetry. In other words, the theory modified 
by $S_M$ is equivalent to actual QCD in the ultraviolet. 

Of course, the addition of (\ref{eq:BRST-breaking}) leads to changes of the theory in the infrared with respect to 
actual QCD and one might ask to what extent such a theory can correctly describe the infrared physics of QCD. 
Interestingly enough, BRST soft breaking terms can be dynamically generated. For example, gluon condensation  
may be incorporated in the action and serve as a starting point for an effective theory containing nonperturbative 
physics \cite{Baulieu:2008fy}. Indeed, the presence of such condensates leads to a nontrivial infrared behavior of 
gluon and ghost propagators, which can be also observed in lattice simulations \cite{Cucchieri:2007rg,Cucchieri:2008fc,Cucchieri:2011ig}. 
In the quark sector of the theory, the BRST breaking term (\ref{eq:BRST-breaking}) is conjectured to be a consequence 
of the condensation of the $\bar qq$ operator. 

The physical meaning of the action $S_{tot}$ becomes clearer after integration in the auxiliary fields $(\xi,\bar\xi)$, 
$(\eta,\bar\eta)$, $(\theta,\bar\theta)$, 
and $(\lambda,\bar\lambda)$. As a result, due to the BRST breaking term $S_M$, one finds a nonlocal action 
\begin{equation}\label{eq:non-local-action}
 S_{nl}=\int d^4x\,\bar\psi_\alpha^i\left[i(\gamma_\mu)_{\alpha\beta}\delta^{ij}\partial_\mu - 
           \delta^{ij}\delta_{\alpha\beta}\left(\frac{2M_1^2M_2}{-\partial^2+m^2} + m_0 \right)\right]\psi^j_\beta
\end{equation}
for the quark sector, plus the minimal coupling to the gauge sector. As a first approximation in our model, we neglect 
the interaction between the gauge and the matter sectors, except through the nonperturbative term $S_M$. Notice that 
the usual local free fermion case can be recovered for $2M_1^2M_2\equiv M_3\rightarrow 0$.

  In the vacuum, the tree level fermion propagator reads (with omission of color and Dirac indices)
\begin{equation}\label{eq:quark-propagator}
 S(p) = -\frac{\gamma_\mu p_\mu + M_0(p)}{p^2+M_0^2(p)},
\end{equation}
where we defined the vacuum mass function
\begin{equation}\label{eq:vacuum-mass-function}
 M_0(p)=\frac{2M_1^2M_2}{p^2+m^2} + m_0.
\end{equation}
Such a quark mass function is compatible with lattice QCD calculations at zero temperature
\cite{Parappilly:2005ei} as well as with an analysis of the Dyson-Schwinger equations of the quark propagator \cite{Bhagwat:2003vw}. 
The parameters $M_3\equiv2M_1^2M_2$, $m^2$ and $m_0$ of the mass function (\ref{eq:vacuum-mass-function}) can be well fitted to the 
lattice results of \cite{Parappilly:2005ei}, giving $M_3=0.196$GeV$^3$, $m^2=0.639$GeV$^2$, and $m_0=0.014$GeV with $\chi^2/dof=1.18$ 
\cite{Dudal:2013vha}. Notice also that the propagator (\ref{eq:quark-propagator}) has two complex conjugate poles and a real pole. 
This can be interpreted as a sign of confinement, a property expected from strongly interacting particles at zero temperature.

Once the action of the quark sector
\begin{equation}\label{eq:local-action}
 S_q = \int d^4x\,\bar\psi_\alpha^i[i(\gamma_\mu)_{\alpha\beta}\partial_\mu - \delta_{\alpha\beta}m_0]\psi_\beta^i + S_{\xi\lambda} + S_M.
\end{equation}
is quadratic in the fermion fields, it is possible to calculate the partition function exactly for arbitrary temperatures 
and chemical potentials. In the next section, we will calculate the grand canonical partition function of the theory defined by 
(\ref{eq:local-action}).

%%%%%%%%%%%%%%%%%%%

\section{The partition function : an exact computation}

In order to calculate the partition function of the theory (\ref{eq:local-action}), we shall use standard techniques of finite temperature 
field theory in the Matsubara imaginary time formalism \cite{LeBellac:book,KapustaGale}. At finite temperature, the 0-direction is compactified 
($\mathbb{R}\rightarrow \mathbb{S}^1\equiv[0,\beta)$), with $\beta\equiv1/T$ and the fermion fields obey anti-periodic boundary 
conditions in imaginary time. As a consequence, the fermion fields can be written in terms of their Fourier transforms as
\begin{equation}\label{eq:Fourier-psi}
 \psi(x_4,{\bf x})= \sum_n\int\,\frac{d^3p}{(2\pi)^3}e^{-i{\bf p}\cdot{\bf x}}e^{-i\omega_n x_4}\tilde\psi(\omega_n,{\bf p}),
\end{equation}
and
\begin{equation}\label{eq:Fourier-psibar}
 \bar\psi(x_4,{\bf x})= \sum_m\int\,\frac{d^3q}{(2\pi)^3}e^{-i{\bf q}\cdot{\bf x}}e^{-i\omega_m x_4}\tilde{\bar\psi}(\omega_m,{\bf q}),
\end{equation}
where $\omega_n=(2n+1)\pi T$ are Matsubara frequencies ($n\in \mathbb{Z}$).

  The presence of a net quark charge requires a nonzero quark chemical potential $\mu$. 
Indeed, a Noether charge can be associated with a global $U(1)$ transformation of the type
\begin{equation}\label{eq:global-U1}
 \psi \rightarrow e^{-i\alpha}\psi,
\end{equation}
with $\alpha\in\mathbb{R}$. It is crucial to notice that the action (\ref{eq:local-action}) is 
invariant under this transformation only if all the auxiliary fields also carry this same charge. 
The corresponding Noether current is then used to impose the grand canonical constraint on the 
hamiltonian of the theory. The calculation is straightforward but it must be carefully performed. 
The details of the derivation of the in-medium effective action for the quark fields are given in 
the Appendix (\ref{app:effective-action}).
 
After imposing the grand canonical constraint and integrating out the auxiliary fields and their 
respective momenta, the grand canonical partition function can then be written in the nonlocal form 
\begin{eqnarray}
 Z(T,\mu) &=& \Tr\exp\left[-\beta\left(\hat H - \mu\hat N\right)\right] \\\nonumber
 &=& \int [D\bar\psi][D\psi]\exp\left[-\int_0^\beta d^4x\, {\cal L}_{nl}[\bar\psi,\psi]\right]
\end{eqnarray}
where, omitting color and spinor indices,
\begin{equation}
 {\cal L}_{nl} = \bar\psi\left[\gamma_4\left(\partial_4-\mu \right) - i{\bf\gamma}\cdot\nabla + \frac{M_3}{-(\partial_4 - \mu)^2 - \nabla^2+m^2} + m_0 \right]\psi.
\end{equation}

That is,  our calculation explicitly showed that  the inclusion of the chemical potential simply amounts 
to the shift $\partial_4\rightarrow \partial_4 - \mu$. It follows from (\ref{eq:Fourier-psi}) and (\ref{eq:Fourier-psibar}) that 
\begin{equation}
 \int_X {\cal L}_{nl} = \sum_n\int\,\frac{d^3p}{(2\pi)^3}{\tilde{\bar\psi}}(-\omega_n,-{\bf p})
 \left[\beta\left(-\gamma_4(i\omega_n+\mu) -  {\bf \gamma}\cdot{\bf p} + M_{n,{\bf p}}(\mu)\right)\right]\tilde\psi(\omega_n,{\bf p})
\end{equation}
where
\begin{equation}\label{eq:mass_function_finiteT}
 \Mp(\mu):=\frac{M_3}{-(i\omega_n+\mu)^2+{\bf p}^2+m^2} + m_0.
\end{equation}
corresponds to the mass function (\ref{eq:vacuum-mass-function}) at finite temperature and chemical potential.

Being quadratic in the fermionic operators, the partition function can be formally integrated immediately,
giving
\begin{equation}\label{eq:partition-function}
 Z(T,\mu)=\det_{p,D,f,c}\left[-\beta\gamma_4(i\omega_n+\mu) - \beta{\bf\gamma}\cdot{\bf p} + \beta M_{n,{\bf p}}(\mu)\right],
\end{equation}
where the (full) determinant has to be taken with respect to the momentum (p), Dirac (D), flavor (f), and color (c) 
subspaces.

Let us follow the same steps as in, e.g., \cite{KapustaGale}, in order to calculate the grand partition function 
(\ref{eq:partition-function}). The determinant in flavor and color spaces are very simple, once all $N_f$ 
flavors and $N_c$ quark colors are on equal footing. This corresponds to taking the determinant in Dirac and 
momentum subspaces to the power $N_cN_f$. Furthermore, the Dirac determinant can be straightforwardly computed 
to be
\begin{equation}
\det_D\left[-\beta\gamma_4(i\omega_n+\mu) - \beta{\bf\gamma}\cdot{\bf p} + \beta M_{n,{\bf p}}(\mu)\right] = 
   \beta^4\left[{\bf p}^2 + \Mp^2(\mu) - (i\omega_n+\mu)^2\right]^2.
\end{equation}

Before proceeding to the calculation of the determinant in momentum space, it is convenient to consider the logarithm of 
the partition function and use the operator identity $\log\det\hat A = \Tr\log\hat A$ where,  in momentum space,  
$\hat A\equiv \beta^4\left[{\bf p}^2 + \Mp^2(\mu) + (\omega_n-i\mu)^2\right]^2$. The trace 
in momentum space correponds to summing over all Matsubara frequencies and all momenta. In the thermodynamic limit 
($V\rightarrow\infty$),
\begin{equation}
 \sum_{\bf p}(\cdots) \longrightarrow V\int\frac{d^3p}{(2\pi)^3}(\cdots)
\end{equation}
As a result, we find
\begin{equation}\label{eq:partition_func_complex_sum}
 \log Z(T,\mu) = 2VN_cN_f\sum_n\int\frac{d^3p}{(2\pi)^3}\log\beta^2\left[{\bf p}^2 + \Mp^2(\mu) - (i\omega_n+\mu)^2\right].
\end{equation}
Notice that the sum (\ref{eq:partition_func_complex_sum}) is real although its integrand is not. Using that 
$\omega_0=-\omega_{-1}$, $\omega_1=-\omega_{-2}$ etc., it is not difficult to write (\ref{eq:partition_func_complex_sum}) 
explicitly as a sum of real numbers by splitting the sum of Matsubara frequencies for $n\geq0$ and $n\leq-1$. In order to 
continue the calculation, it is best to split Eq. (\ref{eq:partition_func_complex_sum}) in two factors, adding 
and subtracting the $\mu=0$ contribution, so that  
\begin{eqnarray}\label{eq:logZ_T0_and_mu}
 \frac{\log Z(T,\mu)}{2\beta VN_cN_f} &=& \sumint\log\left\{\beta^2\left[{\bf p}^2 + \Mp^2(0) +\omega_n^2\right]\right\} + \sumint\log\left\{\frac{{\bf p}^2 + \Mp^2(\mu) - (i\omega_n+\mu)^2}{{\bf p}^2 + \Mp^2(0) +\omega_n^2}\right\}\crcr
                            &\equiv&  \frac{\log Z(T,0)}{2\beta VN_cN_f} +  \frac{\log Z^{(\mu)}(T,\mu)}{2\beta VN_cN_f},
\end{eqnarray}
  where we used the standard sum-integral notation,
\begin{equation}
 \sumint (\cdots) \equiv T \sum_{n=-\infty}^\infty\int\frac{d^3p}{(2\pi)^3}(\cdots).
\end{equation}

One interesting feature of the decomposition (\ref{eq:logZ_T0_and_mu}) is that only the term $\log Z(T,0)$ contains any divergencies. 
They are the same as those of the local theory at $T=0$ and therefore can be straightforwardly subtracted. The $\mu\not=0$ contribution 
is finite, as expected from the argument that the introduction of a chemical potential should not bring any new divergencies to the theory. 
For this reason, it can be calculated numerically and also has a closed expression in terms of elementary integrals in the limit of zero 
temperature.

For clarity, let us now separately discuss the calculations of the $\mu=0$ and $\mu\not=0$ contributions.

%%%
\subsection{The $\mu=0$ contribution}

In the standard calculation of the partition function of local free quarks of mass $m_0$ at $\mu=0$ 
(see, e.g., \cite{LeBellac:book,KapustaGale}), one is lead to the calculation of the sum-integral
\begin{equation}\label{eq:logZ_local}
  \log Z_{loc}(T,0) = 2N_cN_f \beta V\sumint \log \left\{\beta^2 \left[\omega_n^2+\omega^2 \right]\right\},
\end{equation}
where $\omega=\sqrt{{\bf p}^2+m_0^2}$. After subtractions of $T$- and $\mu$-independent (infinite) constants, one arrives at the final expression 
\begin{equation}
 \log Z_{loc}(T,0) = 2N_cN_f \beta V\int\frac{d^3p}{(2\pi)^3}\left[\omega + 2T\log\left(1+e^{-\beta\omega}\right)\right].
\end{equation}

It is quite straightforward to show that the calculation of 
\begin{equation}\label{eq:logZ_zero-mu}
 \log Z(T,0) = 2\beta VN_cN_f  \sumint\log\left\{\beta^2\left[{\bf p}^2 + \Mp^2(0) +\omega_n^2\right]\right\} 
\end{equation}
can be reduced to the sum of four terms which are each formally identical to (\ref{eq:logZ_local}). Indeed, using 
Eq. (\ref{eq:mass_function_finiteT}), it is possible to write the argument of the logarithm in (\ref{eq:logZ_zero-mu}), 
$Arg$, as a ratio of two polynomials, 
\begin{equation}\label{eq:arg-log}
 Arg:=\frac{\beta^2
 \left[({\bf p}^2 + m_0^2 + \omega_n^2)({\bf p}^2 + m^2 + \omega_n^2)^2 + 2M_3m_0({\bf p}^2 + m^2 + \omega_n^2) + M_3^2\right]}{({\bf p}^2+m^2+\omega_n^2)^2} 
 \equiv \beta^2\frac{P_3(\omega_n^2)}{({\bf p}^2+m^2+\omega_n^2)^2}         
\end{equation}
where the third degree polynomial $P_3(\omega_n^2)$ can be factored out as
\begin{equation}
 P_3(\omega_n^2) = (\omega_n^2 + \varphi_1^2)(\omega_n^2 + \varphi_2^2)(\omega_n^2 + \varphi_3^2),
\end{equation}
with $-\varphi_i^2$ $(i=1,2,3)$ the three roots of $P_3$. In general, the roots of a third degree polynomial of real 
coefficients are one real number and a couple of complex conjugate numbers. This is indeed the case for our parameter 
set, as should be evident from the presence of complex poles in the propagator, i.e., complex masses. The three roots 
can be explicitly calculated as functions of ${\bf p}^2$ and the parameters $M_3$, $m^2$, and $m_0$. Substituting back 
in (\ref{eq:logZ_zero-mu}), we have
\begin{equation}
 \log Z(T,0) = 2N_cN_f \beta V\sumint \left\{\sum_{i=1}^3\log\left[\beta^2\left(\omega_n^2+\varphi_i^2\right) \right] - 2\log\left[\beta^2\left(\omega_n^2+{\bf p}^2+m^2\right)\right] \right\}.
\end{equation}
As advertised,   each term   has the same structure as (\ref{eq:logZ_local}). Therefore, it follows straightforwardly that
\begin{equation}\label{eq:log-Z_zero_mu-final}
 \log Z(T,0) = \log Z_0 + 4N_cN_f V\int\frac{d^3p}{(2\pi)^3}
      \log\left[\frac{\left(1+e^{-\beta\varphi_1}\right)\left(1+e^{-\beta\varphi_2}\right)\left(1+e^{-\beta\varphi_3}\right)}{\left(1+e^{-\beta\varphi_0}\right)^2}\right],
\end{equation}
where
\begin{equation}
 \log Z_0 = 2N_cN_f \beta V\int\frac{d^3p}{(2\pi)^3}\left(\varphi_1+\varphi_2+\varphi_3-2\varphi_0\right) 
\end{equation}
is the pure vacuum contribution and $\varphi_0=\sqrt{{\bf p}^2 + m^2}$. We choose a normalization such that $\log Z(0,0)\equiv0$, i.e., $\log Z_0=0$. Finally, 
notice that although two of the $\varphi_i$ are complex, their imaginary parts cancel in (\ref{eq:log-Z_zero_mu-final}), so that the final result is real, 
as it had to be.

%%%%%%%%
\subsection{The finite $\mu$ contribution and its zero temperature limit}

The $\mu\not=0$ contribution in (\ref{eq:logZ_T0_and_mu}),
\begin{equation}\label{eq:logZ_finite-mu}
 \log Z^{(\mu)}(T,\mu) = 2\beta VN_cN_f\sumint\log\left\{\frac{{\bf p}^2 + M_{n,{\bf p}}^2(\mu) - (i\omega_n+\mu)^2}{{\bf p}^2 + M_{n,{\bf p}}^2(0) +\omega_n^2}\right\},
\end{equation}
for the partition function is finite and can be calculated 
numerically for any finite $T$ and $\mu$. Unfortunately, we could not find a closed expression for it, except in 
the interesting $T\rightarrow0$ limit, where it equals the full partition function. The zero temperature limit is 
taken most easily by expressing the Matsubara sum in (\ref{eq:logZ_finite-mu}) as a contour integral in the complex 
plane \cite{KapustaGale}. It is adequate to choose the rectangular integration path as 
${\cal P}=(-iL+\mu+\epsilon,+iL+\mu+\epsilon) \cup (+iL+\mu+\epsilon,+iL+\mu-\epsilon) \cup (+iL+\mu-\epsilon,-iL+\mu-\epsilon) 
\cup (-iL+\mu-\epsilon,-iL+\mu+\epsilon)$, with $L\rightarrow\infty$ and $\epsilon\rightarrow0^+$. Defining
\begin{equation}\label{eq:nonlocal_dispersion_relation}
\Omega_{{\bf p}}^2(\zeta):={\bf p}^2 + \left[\frac{M_3}{-\zeta+{\bf p}^2 + m^2} + m_0 \right]^2
\end{equation}
and
\begin{equation}
 f(\xi) := \log\left\{\frac{\Omega_{{\bf p}}^2(\xi^2) - \xi^2}{\Omega_{{\bf p}}^2[(\xi-\mu)^2] - (\xi-\mu)^2}\right\},
\end{equation}
we can write the sum over Matsubara frequencies in (\ref{eq:logZ_finite-mu}) as

\begin{eqnarray}\label{eq:Matsubara_sum_1}
 T\sum_n f(i\omega_n+\mu) = &&\!\!\!\!\! T \oint_{\cal P}\frac{d\xi}{2\pi i}f(\xi)\frac{\beta}{2}\tanh\left(\frac{\beta (\xi-\mu)}{2}\right) \crcr
                      = &&\!\!\!\!\! \int_{-i\infty+\mu+\epsilon}^{i\infty +\mu + \epsilon}\frac{d\xi}{4\pi i}\,f(\xi)\tanh\left[\frac{\beta (\xi-\mu)}{2}\right] 
                      + \int_{i\infty+\mu-\epsilon}^{-i\infty +\mu - \epsilon}\frac{d\xi}{4\pi i}\,f(\xi)\tanh\left[\frac{\beta (\xi-\mu)}{2}\right].
\end{eqnarray}
In the $T\rightarrow0$ ($\beta\rightarrow\infty$) limit, 
\begin{equation}
 \tanh\left[\frac{\beta(\xi-\mu)}{2}\right]\rightarrow 1-2\theta\left[{\rm Re}(\mu - \xi)\right],
\end{equation}
so that it follows from (\ref{eq:Matsubara_sum_1}) that 
\begin{eqnarray}\label{eq:Matsubara_sum_2}
 T\sum_n f(i\omega_n+\mu) = &&\!\!\!\!\! \int_{-i\infty+\mu+\epsilon}^{i\infty +\mu + \epsilon}\frac{d\xi}{4\pi i}\,f(\xi)
                      - \int_{i\infty+\mu-\epsilon}^{-i\infty +\mu - \epsilon}\frac{d\xi}{4\pi i}\,f(\xi) \crcr
                      = &&\!\!\!\!\! \int_{-i\infty+\mu}^{i\infty +\mu}\frac{d\xi}{2\pi i}\,f(\xi)\crcr
                      = &&\!\!\!\!\! \int_{-\infty}^{\infty}\frac{d\theta}{2\pi}\,f(i\theta + \mu)
\end{eqnarray}
The analiticity of $f$ was also used to drop the $\epsilon$ term in the integration limits and, in the last step, we made the 
change of variables $\xi\rightarrow\theta=-i(\xi-\mu)$.  The zero-temperature limit of the complete partition function is thus
\begin{eqnarray}\label{eq:logZ_finite-mu_zeroT}
  \log Z(0,\mu) &=& \log Z^{(\mu)}(0,\mu) = 2\beta VN_cN_f\int\frac{d^3p}{(2\pi)^3}\int_{-\infty}^{\infty}\frac{d\theta}{2\pi}f(i\theta+\mu)\crcr
                &=& 2\beta VN_cN_f\int\frac{d^3p}{(2\pi)^3}\int_{0}^{\infty}\frac{d\theta}{2\pi}\left[f(i\theta+\mu)+f(-i\theta+\mu)\right].
\end{eqnarray}
It is important to notice that once $f(-i\theta+\mu)=f^*(i\theta+\mu)$, $\forall \theta\in\mathbb{R}$, 
the partition function (\ref{eq:logZ_finite-mu_zeroT}) is a real quantity, as it had to be.

As a crosscheck, it is possible to show that for the local case, $M_3=0$, one simply has 
$\tilde\Omega^2_{p_0,{\bf p}}(\mu)={\bf p}^2 + m_0^2$ and thus
\begin{equation}\label{eq:logZ_zeroT_local}
 \log Z_{loc}(0,\mu) = 2\beta V N_cN_f \int\frac{d^3p}{(2\pi)^3}\left[\mu - \sqrt{{\bf p}^2 + m_0^2}\right]
 \Theta\left(\mu - \sqrt{{\bf p}^2 + m_0^2}\right),
\end{equation}
which is the well known result for free local quarks \cite{LeBellac:book,KapustaGale}. 

This ends our calculation of the partition function of the model at finite temperature and quark chemical potential. Notice that no approximations were 
used after the introduction of the model. In the next section, we will present our results for thermodynamics quantities directly calculated from the 
partition function $Z(T,\mu)$.

%%%%%%%%%%%%%%%%%%%%%%%%%%%%%%%%%%%%%%%%%%%%%%%%%%%%%%%%%%%%%%
\section{Thermodynamic quantities}

Starting from the partition function of the model,
\begin{equation}\label{eq:partition_function_full}
 \frac{\log Z(T,\mu)}{2\beta VN_cN_f} = 2T \int\frac{d^3p}{(2\pi)^3}
      \log\left[\frac{\left(1+e^{-\beta\varphi_1}\right)\left(1+e^{-\beta\varphi_2}\right)\left(1+e^{-\beta\varphi_3}\right)}{\left(1+e^{-\beta\varphi_0}\right)^2}\right]  
      + \sumint\log\left\{\frac{\Omega_{n,{\bf p}}^2(\mu) +(\omega_n-i\mu)^2}{\Omega_{n,{\bf p}}^2(0) +\omega_n^2}\right\},
\end{equation}
it is possible to calculate any thermodynamic property of the system. For example, the pressure is given by 
\begin{equation}
 P(T,\mu) = \frac{T}{V}\log Z(T,\mu),
\end{equation}
which equals\footnote{As long as the normalization condition $\log Z(0,0)=0$ holds.} minus the thermodynamic potential 
energy per unit volume, $\Omega(T,\mu)=-P(T,\mu)$. Therefore, other 
quantities may be written as derivatives of the pressure, such as entropy density,
\begin{equation}
 s(T,\mu) = -\frac{\partial \Omega}{\partial T}(T,\mu) = \frac{\partial P}{\partial T}(T,\mu)
\end{equation}
or the quark number density
\begin{equation}\label{eq:quark_density}
 n(T,\mu) = -\frac{\partial \Omega}{\partial \mu}(T,\mu) = \frac{\partial P}{\partial \mu}(T,\mu).
\end{equation}
The energy density can be derived from the other quantities using the thermodynamic identity $e=Ts-P+\mu n$, while 
the quark number susceptibility from the second derivative of the pressure,
\begin{equation}\label{eq:quark_susceptibility}
 \chi(T,\mu) = -\frac{\partial^2\Omega}{\partial\mu^2}(T,\mu) = \frac{\partial n}{\partial \mu}(T,\mu)
\end{equation}

 In what follows we analyze different regimes of the thermodynamics of the nonlocal quark model. We recall that all quantities are calculated from 
the partition function (\ref{eq:partition_function_full}) with $N_c=3$ and $N_f=2$.

\subsection{ Thermal case}

 Let us first address the thermodynamics of a medium of hot confined quarks at zero chemical potential. In this regime lattice simulations provide a robust reference for the full QCD case.

%%%%%%%%%%%%%%%%%%%%
\begin{figure}[h!]
 \centering
 \subfloat[]{\includegraphics[width=7.8cm]{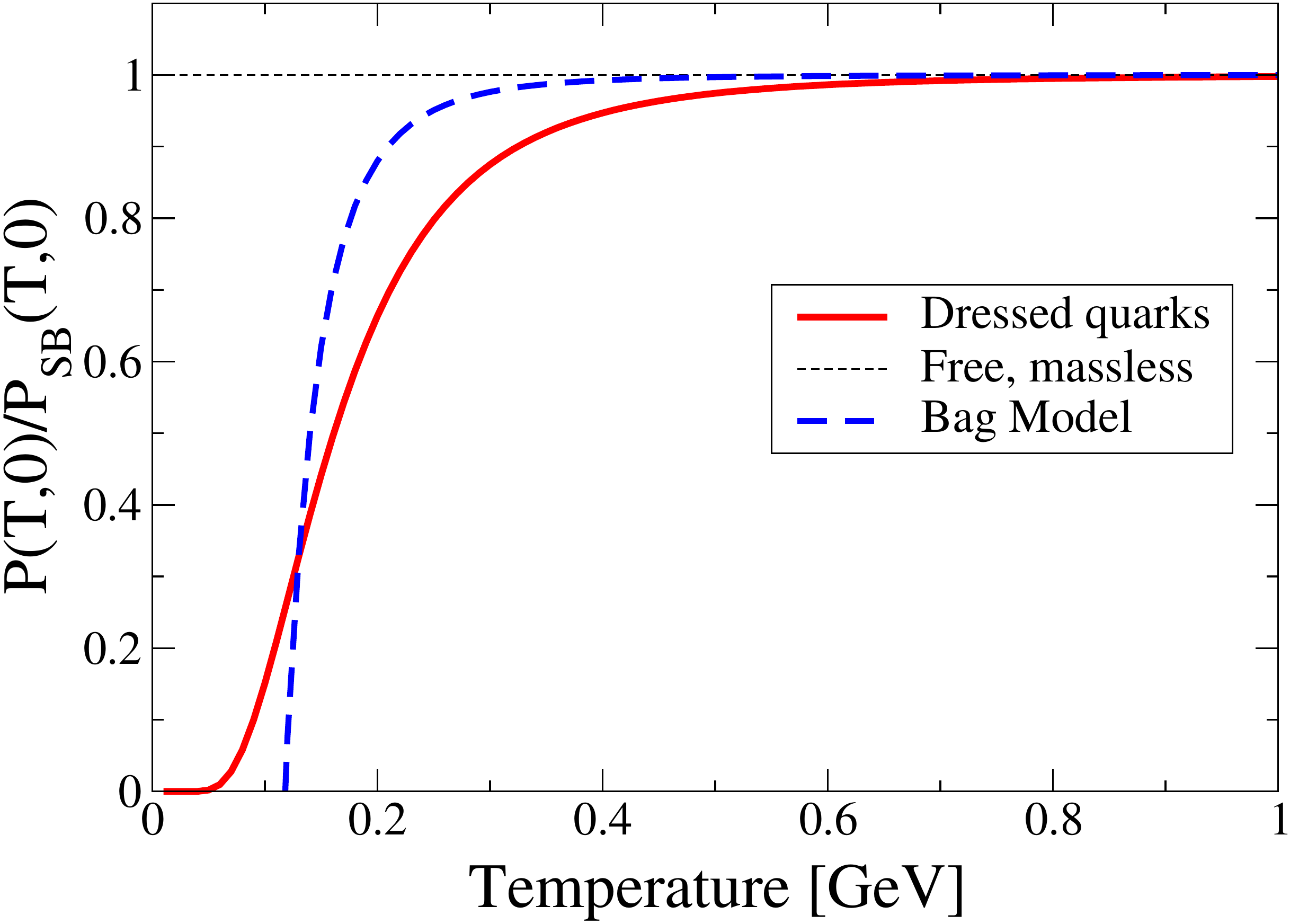}}
 \subfloat[]{\includegraphics[width=8.6cm]{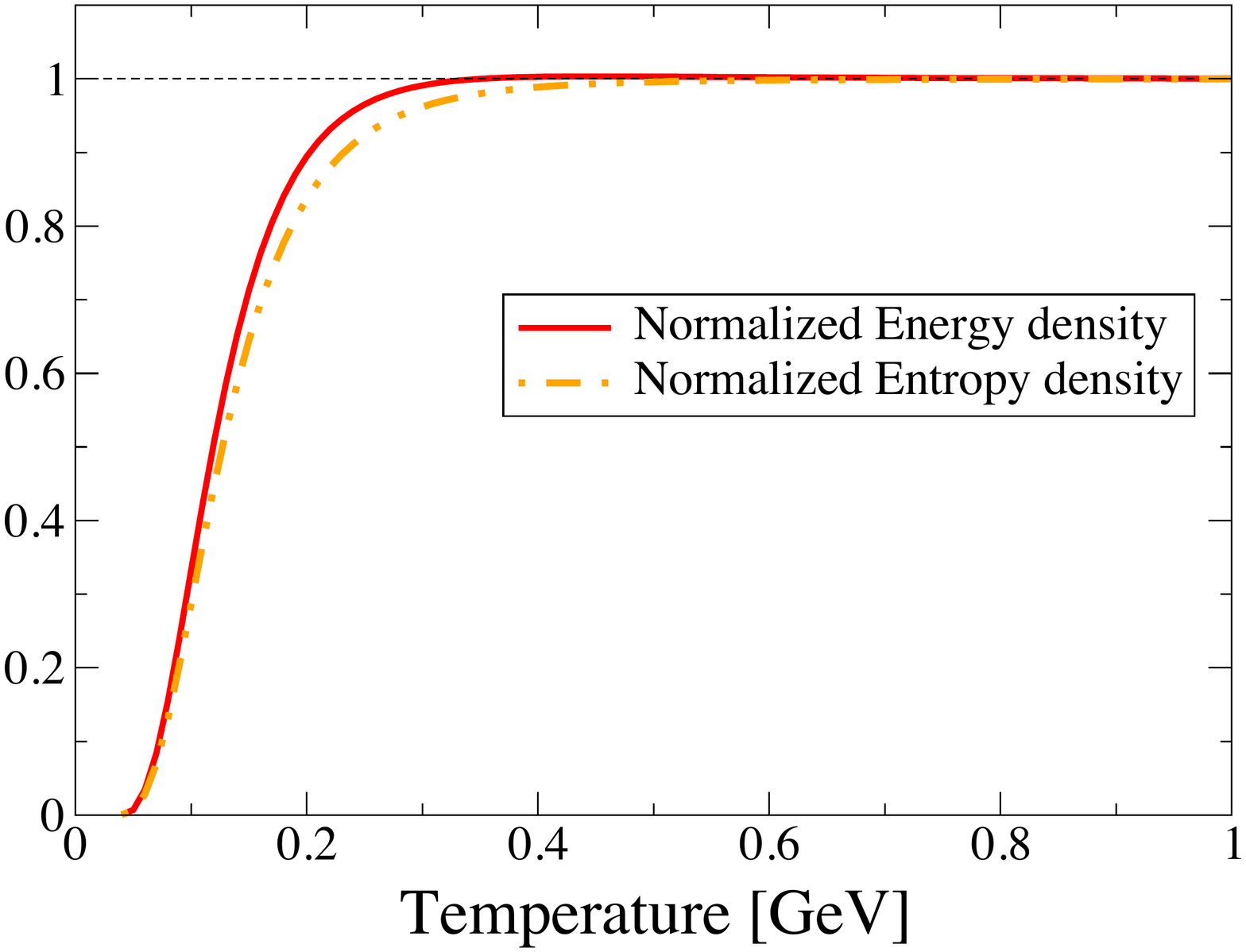}}
 \caption{(a) Pressure as a function of temperature for $\mu=0$, normalized by the free gas limit 
 (\ref{eq:stefan-boltzmann-pressure}) for nonlocal quarks and for the MIT bag model with bag 
 constant $B=(0.145{\rm GeV})^4$. (b) Energy and entropy densities, normalized by the respective 
 free gas limits.}
 \label{fig:pressure-energy-entropy-at-zero-mu}
\end{figure}
%%%%%%%%%%%%%%%%%%%%

In Fig. (\ref{fig:pressure-energy-entropy-at-zero-mu}a), we show the pressure as a function of the temperature at 
zero chemical potential. We normalize the result by the pressure of free, massless fermions,
\begin{equation}\label{eq:stefan-boltzmann-pressure}
 P_{SB}(T,\mu) = N_cN_f\left[\frac{7\pi^2 T^4}{180} + \frac{\mu^2 T^2}{6} + \frac{\mu^4}{12\pi^2}\right],
\end{equation}
 evaluated at $\mu=0$. For comparison, we  also display  the pressure of massless quarks subject to a constant negative {\it bag pressure} 
$p_{Bag}=-B=-(0.145{\rm GeV})^4$. 

A smooth but fast rise of the pressure as temperature increases is observed, indicating a drastic change in the number of thermal degrees of freedom between low- and high-temperature systems. 
A similar behavior is seen for the thermal crossover in lattice QCD simulations.
In contrast to the bag model, our model is capable of describing the whole range of temperatures, with no negative pressures attained at low temperatures. This corroborates the thermodynamical stability of the model and its consistency. The associated energy and entropy densities at $\mu=0$, 
normalized by their respective free gas limits, present a similar behavior as functions of temperature, as can be seen in Fig. (\ref{fig:pressure-energy-entropy-at-zero-mu}b).

\begin{figure}[h!]
\begin{center}
 \centering
 \subfloat[]{\includegraphics[width=7.8cm]{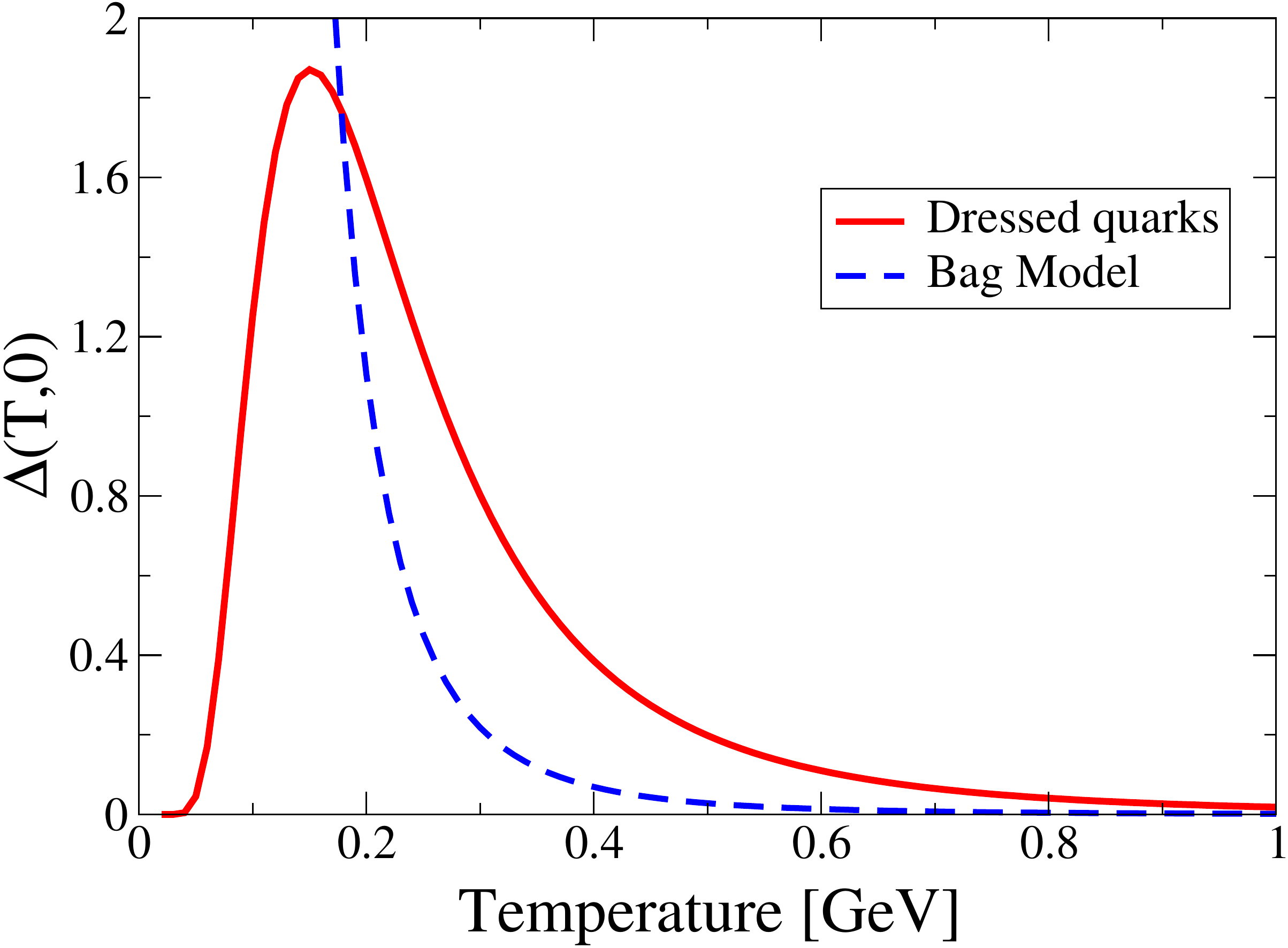}}
 \hspace{1cm}
 \subfloat[]{\includegraphics[width=7.8cm]{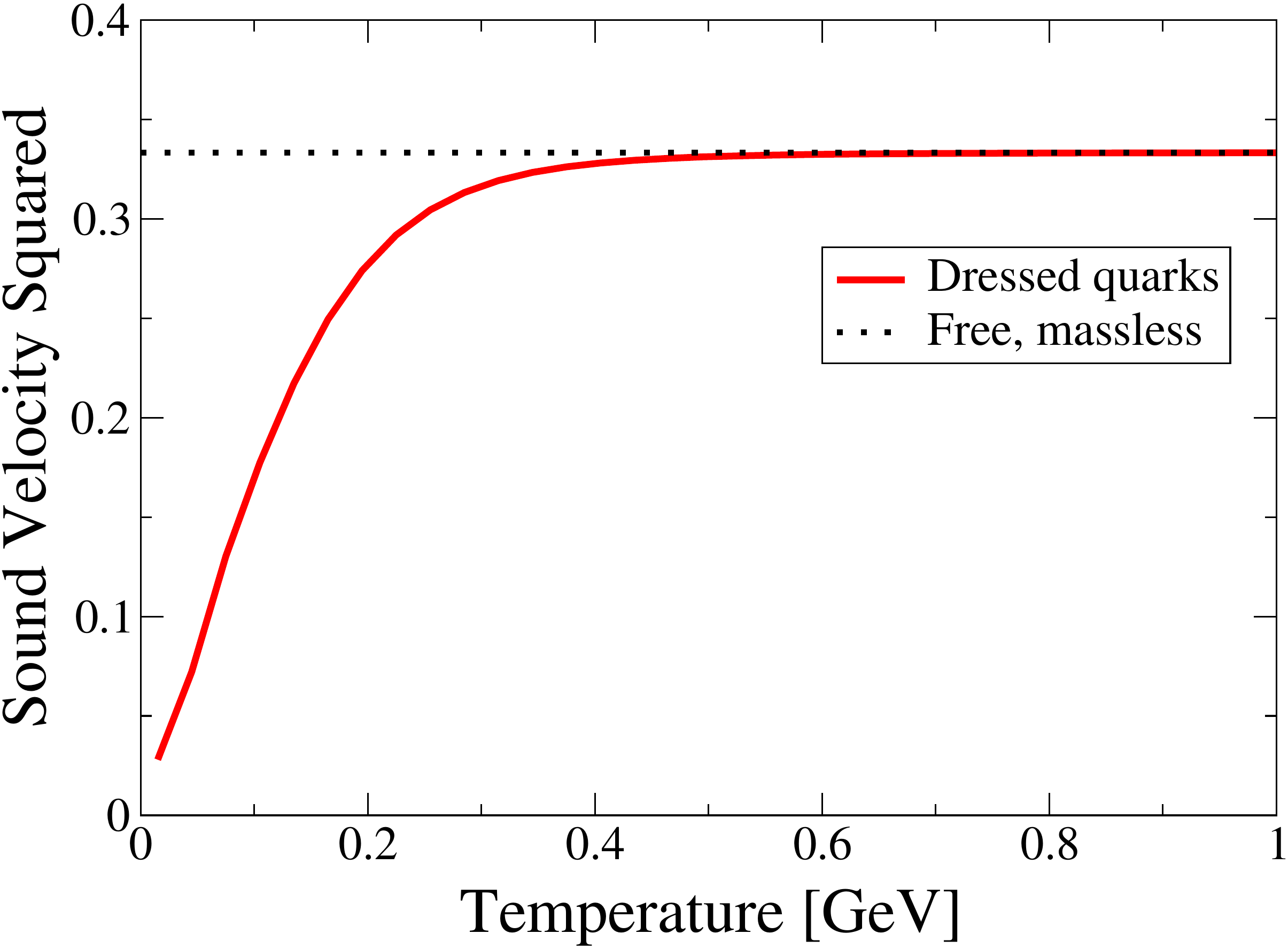}}
\caption{(a) Trace anomaly as a function of temperature for zero chemical potential, normalized by $T^4$. A clear distinction 
from the bag model result is seen. (b) Sound velocity squared as a function of temperature.}
\label{fig:trace-anomaly-and-sound-velocity}
\end{center}
\end{figure}

Another thermodynamic observable of interest is the (normalized) trace anomaly or ``interaction measure''
\begin{equation}
 \Delta(T,\mu) = \frac{E - 3P}{T^4}.
\end{equation}
It corresponds to the deviation from the tracelessness of a conformal energy-momentum tensor 
(normalized by $T^4$), or, in other words, it measures how different the nonlocal quark system 
behaves with respect to an ideal gas of massless quarks. The trace anomaly for the nonlocal quarks, 
plotted in Fig. (\ref{fig:trace-anomaly-and-sound-velocity}a), is clearly different from the bag model result and shows a peak around $T\simeq0.15$GeV. 
This suggests a smooth transition between two quasi conformal phases at low and high temperatures. 
However, in such a transition, the sound velocity squared,
\begin{equation}\label{eq:sound-velocity}
 c_s^2 = \frac{\partial P}{\partial E},
\end{equation}
typically presents a minimum around the transition 
temperature and this is {\it not} seen in (\ref{fig:trace-anomaly-and-sound-velocity}b).

%%%%%%%%%%%%%%%%%%%%%%%%%%%%%
%%%%%%%%%%%%%%%%%%%%%%%%%%%%%

Due to asymptotic freedom and chiral symmetry restoration, encoded in the $p\rightarrow\infty$ limit of the mass 
function (\ref{eq:vacuum-mass-function}), it is expected that the system approaches the limit of free, massless quarks as 
temperature increases. This is indeed seen from our results for all the thermodynamic 
quantities we calculated. On the other hand, the non-trivial interactions included in our model via the nonlocal 
background show up in the results for
intermediate temperatures.
There is a clear change of behavior of all thermodynamic quantities 
at $T\simeq0.15$GeV, the typical temperature scale of the chiral or deconfined phase crossover. In particular, the 
trace anomaly shows a peak precisely at this temperature, with a steep rise from zero as observed e.g. in
lattice QCD simulations and in contrast to QM models. It is interesting that the inclusion of a 
well-chosen Polyakov-loop potential (fitted to thermal lattice data) may provide exactly this steep rise in the trace anomaly at low temperatures, which is absent in pure chiral models \cite{Schaefer:2009ui}. In this sense, and for this particular observable, our nontrivial background -- encoded in the nonlocal quark propagator fitted to zero-temperature lattice results -- seems to play a similar role as the Polyakov-loop potential, further suggesting its connection to confinement and nonperturbative physics.

%%%%%%%%%%%%%%%%%%%%%%%%%%%%%
\begin{figure}[h!]
\begin{center}
 \centering
\includegraphics[width=7.5cm]{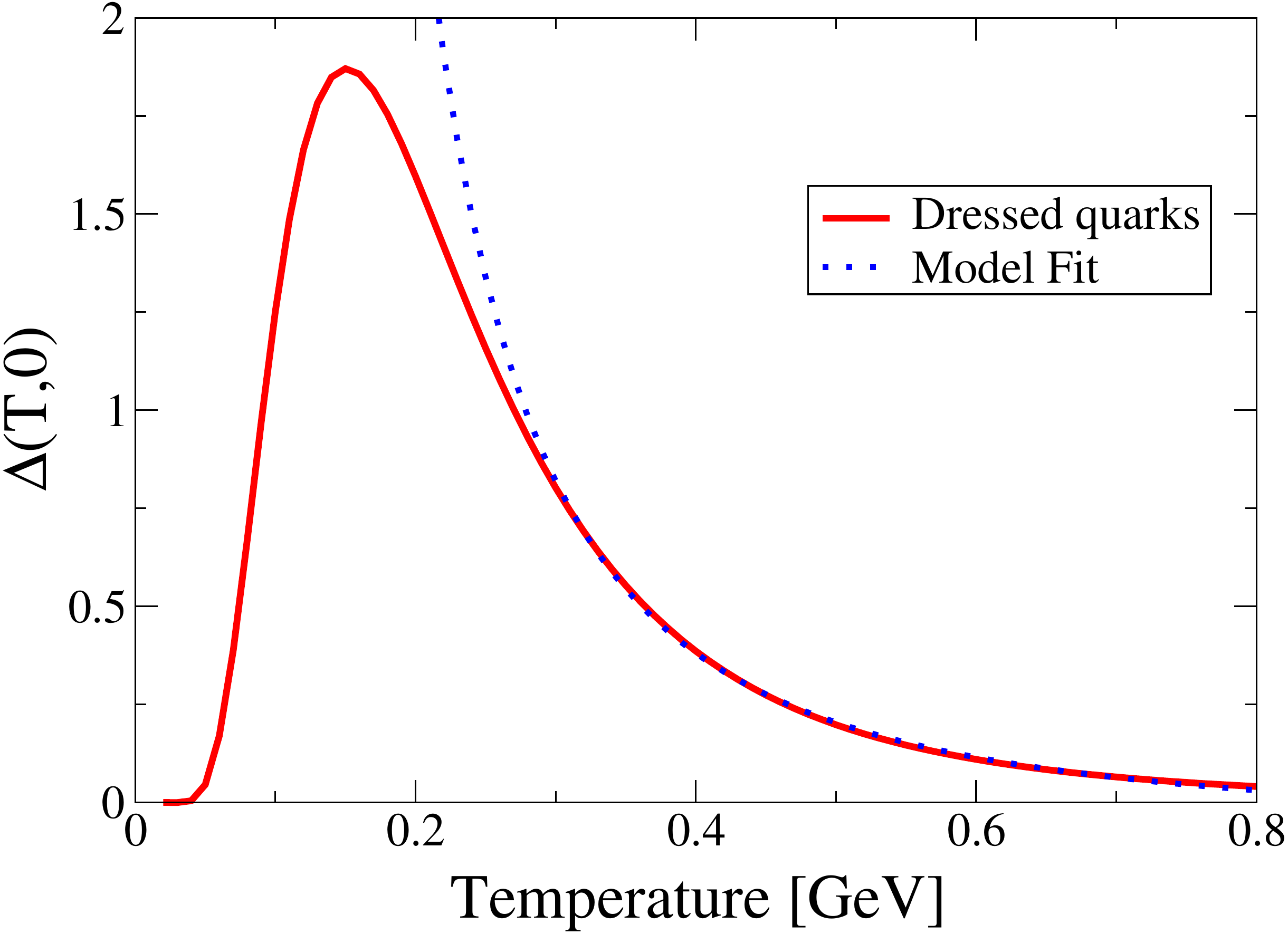}
\caption{High-temperature fit (dotted line) of the trace anomaly as a function of temperature for zero chemical potential, normalized by $T^4$. The solid, red line represent our model results.}
\label{fig:trace-anomaly-fit}
\end{center}
\end{figure}
%%%%%%%%%%%%%%%%%%%%%%%%%%%%%

In order to further understand the physics contained in our result for temperatures above the peak, we display in Fig.\ref{fig:trace-anomaly-fit}  a fit of the trace anomaly in the temperature interval $T\in [300,800]$ MeV in the following form:
\begin{equation}
[\Delta(T,\mu=0) ]_{\rm high\,T}=  a+\frac{b}{T^2}+\frac{c}{T^4}\,,
\label{fit}
\end{equation}
with $a=-0.069$, $b= 0.062$ GeV$^2$ and $c= 0.0016$ GeV$^4$, with the $O(T^{-2})$ term dominant over the $O(T^{-4})$ one in this range.

A fit of the form \eqref{fit} was also performed for lattice data in Ref.\cite{Cheng:2007jq}. There they observe a similar hierarchy of contributions, with $b_{\rm latt}\sim 0.1$ GeV$^2$ and $c_{\rm latt}\sim 0.02$ GeV$^4$. We note, however, that direct comparison of these results should be taken with care, since the system from Ref.\cite{Cheng:2007jq} is (2+1)-flavour full QCD, while our model describes only heavier-than-physical degenerate up and down quarks\footnote{The light quarks in Ref.\cite{Cheng:2007jq} also have larger-than-physical masses, corresponding to $m_{\pi}=220$ MeV, while the pion in the lattice data \cite{Parappilly:2005ei} for the quark propagator fitted here has $m_{\pi}
\sim 240$ MeV.}, with gluons not explicitly included (their influence appears only in the nonperturbative quark dressing). It is nevertheless encouraging that the same structure and hierarchy of contributions is found in our confining quark model, in contrast to the bag model or a free massive quark system, as will be detailed below.

The three terms in the fit \eqref{fit} may be associated with different physical contributions. The constant $a$ is related to a logarithmic dependence in the pressure, i.e. $\sim\log T$,
that could in principle be mapped to the perturbative result (including the running coupling), if the temperature is high enough. The parameter $c$ is originated by a constant pressure, which is negative for positive $c$. This is exactly the type of contribution included in an {\it ad hoc} manner in a bag model to mimic confinement, and that appears here as a natural consequence of the nontrivial background considered. An estimate of the bag constant predicted in the high-temperature region of our model can then be obtained from $c=4B$, yielding $B\sim (141\, {\rm MeV})^4$ which nicely falls in the ballpark of the values adopted in the literature of QCD models.

The $O(T^{-2})$ term in the trace anomaly, however, is usually absent or very small in a na\"ive bag model. 
The smallness of this term is directly related to the lightness of up and down quarks. Indeed, a {\it massive} bag model has a pressure of the form \cite{KapustaGale}:
\begin{equation}
P_{\rm bag}= \#_0 T^4 - \#_1 m^2T^2-B\,,
\end{equation}
where $m$ is the current quark mass. The respective trace anomaly is 
$T^4\Delta_{\rm bag}=4B+\#_2 m^2T^2$. Taking $m=14$ MeV, the coefficient of $T^{-2}$ is $\sim 10^{-4}$ GeV$^2$, three orders of magnitude below the value obtained in this fit of our model and also in lattice QCD. It becomes clear that 
the presence of a dominant $O(T^{-2})$ contribution may be linked to the generation of a large mass scale in the nonperturbative quark sector, as the $M_3$ parameter in our model. It is worthwhile noting that even a bag model that considers effective quark masses large enough to give the desired $T^{-2}$ contribution would fail to reproduce lattice data for $T<300$ MeV, due to the nonperturbative peak structure that is developed for lower temperatures.

%. It is clear that a dominant 
%
%\footnote{
%For pure gauge systems, a $O(T^{-2})$ term is ruled out by dimensional analysis in the absence of dynamically generated mass scales. In this case,  
%} 
%
%It has been associated e.g. with nonperturbative effects from Wilson lines in pure gauge systems \cite{Pisarski}.
%
%A fit of the form \eqref{fit} was also performed for lattice data in Ref.\cite{Cheng:2007jq}. There they observe a similar hierarchy of contributions, with $b_{\rm latt}\sim 0.1$ GeV$^2$ and $c_{\rm latt}\sim 0.02$ GeV$^4$. We note, however, that direct comparison of these results should be taken with care, since the system from Ref.\cite{Cheng:2007jq} is (2+1)-flavour full QCD, while our model describes only heavier-than-physical degenerate up and down quarks\footnote{The light quarks in Ref.\cite{Cheng:2007jq} also have larger-than-physical masses, corresponding to $m_{\pi}=220$ MeV, while the pion in the lattice data for the quark propagator fitted here has $m_{\pi}
%\sim 240$ MeV.}, with gluons not explicitly included (their influence appears only in the nonperturbative quark dressing). It is nevertheless encouraging that the same structure and hierarchy of contributions is found in our confining quark model, in contrast to the bag model or a free massive quark system, as will be detailed below.

\subsection{ Cold and dense confined quarks}

Let us now turn our attention to the $T\rightarrow 0$ case and study some thermodynamic quantities as functions of the 
chemical potential $\mu$.  This regime of QCD presents a severe Sign Problem which hinders
the application of Montecarlo simulations, so that model predictions are  extremely valuable tools to 
explore this region of the phase diagram. 

%%%%%%%%%%%%%%%%%%%%
\begin{figure}[!h]
 \centering
 \subfloat[]{\includegraphics[width=8.5cm]{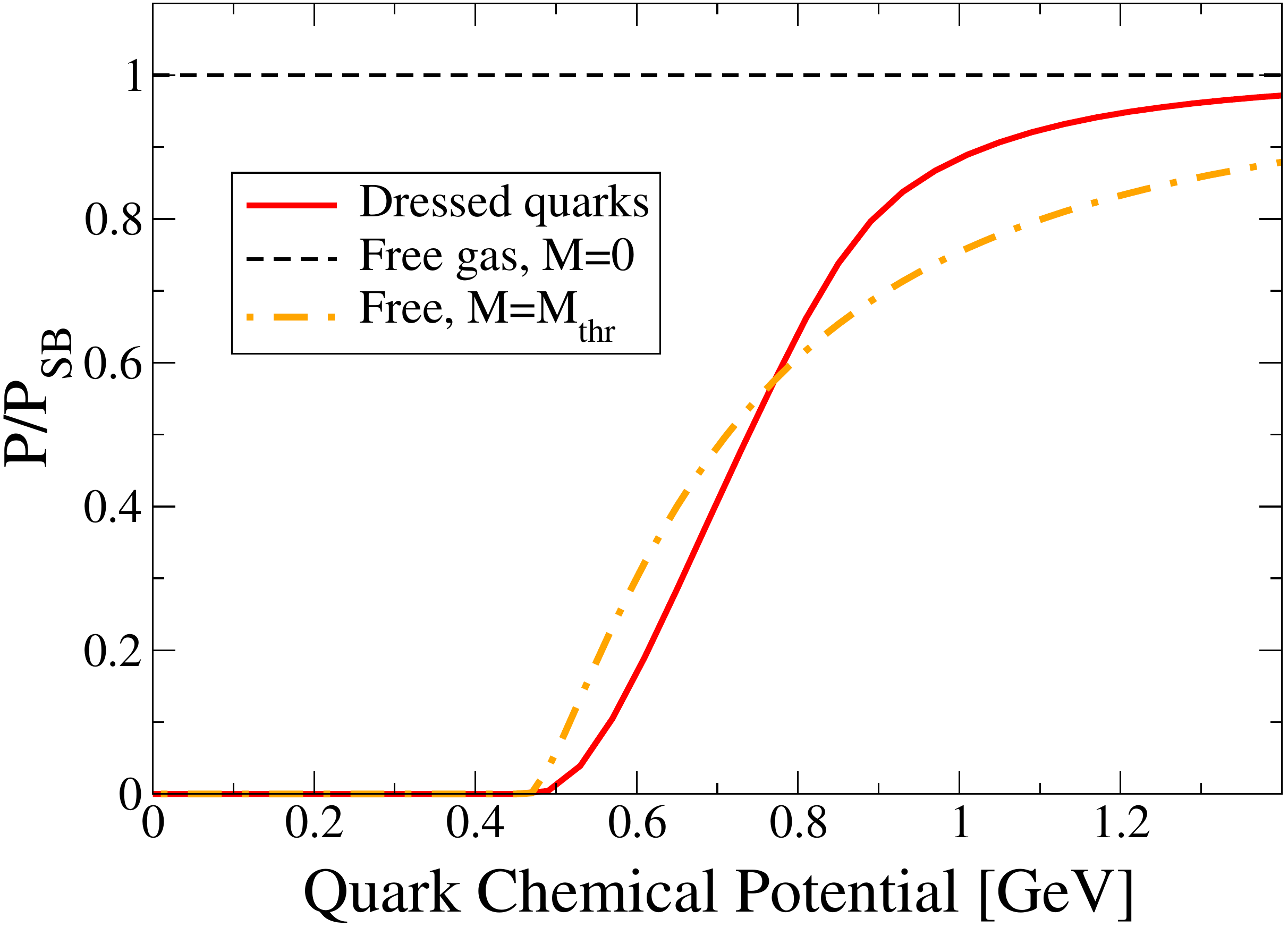}}
 \caption{Pressure as a function of the chemical potential for $T=0$, normalized by the free gas limit 
 $P_{SB}(0,\mu)=N_cN_f\mu^4/12\pi^2$. The massless (dotted) and massive (dot-dashed) free gas results are plotted for comparison. We use $M_{thr}=467$ MeV.}
 \label{fig3:pressure_zero_temperature}
\end{figure}
%%%%%%%%%%%%%%%%%%%%

In Fig. (\ref{fig3:pressure_zero_temperature}), we show the pressure as a function of the chemical potential, 
  at zero temperature. We compare the pressure from the model 
with that of local massive quarks of mass $M_{thr}=0.467$ GeV,  both normalized to the pressure of massless quarks.  The choice of the mass $M_{thr}$ is the one that provides a pressure that better compares to our model results in order to make explicit the differences and the specific features that are exclusive to the confining quark model here studied.  

  Since the system is at zero temperature there is no thermal energy to excite particles at low densities, yielding a vanishing thermodynamical response. As already observed in the thermal case, the results are fully stable and consistent with general physical expectations for a cold and dense system. In particular,  the pressure vanishes for chemical potentials up to some value  $\mu\simeq M_{thr}= 0.467\,$GeV, which is consistent with a dynamically generated scale and not directly connected to any specific mass parameter of the model.  Starting at 
this point, the pressure smoothly rises until reaching the limiting pressure of local quarks for high chemical potentials. It is not 
surprising that the massless limit is  attained  at large chemical potentials, given the $\mu$-dependence of the  
momentum-dependent mass function in Eq. (\ref{eq:mass_function_finiteT}).

%%%%%%%%%%%%%%%%%%%%
\begin{figure}
 \centering
 \subfloat[]{
  \includegraphics[width=8.5cm]{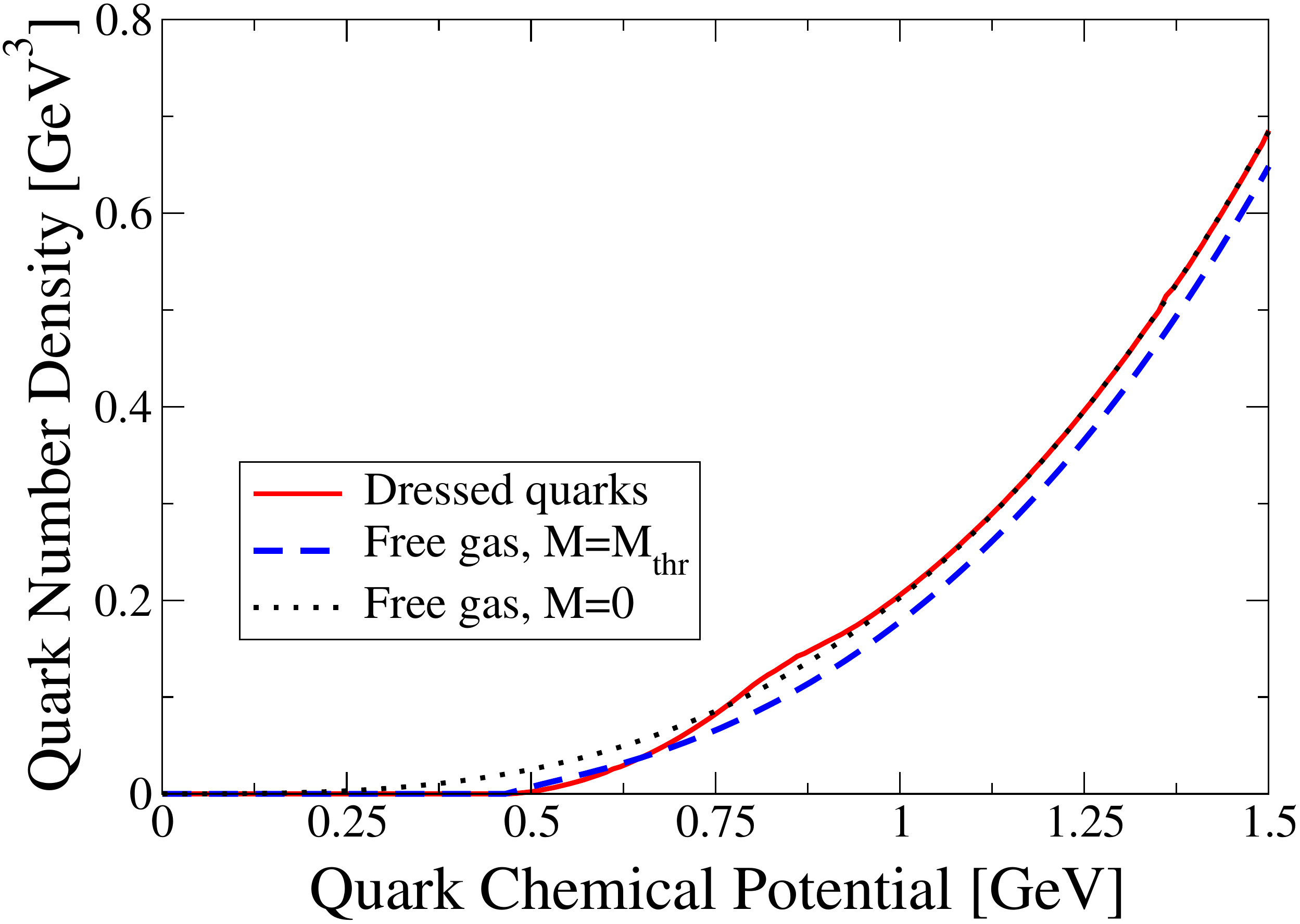}\llap{\makebox[4.cm][l]{\raisebox{3.2cm}{\hspace{-2cm}\includegraphics[height=2.7cm]{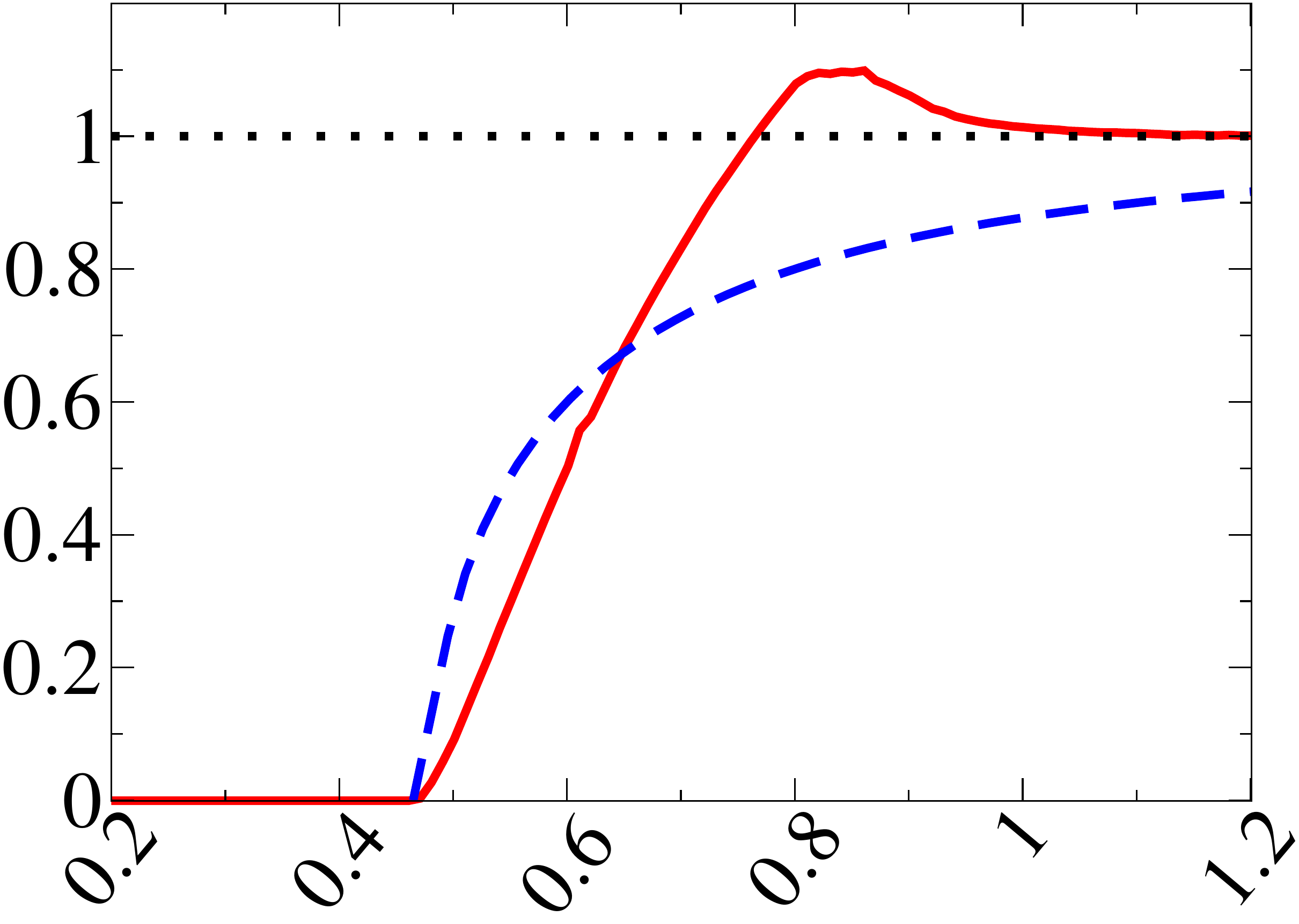}}}} }
\hspace{0.5cm} \subfloat[]{\includegraphics[width=8.5cm]{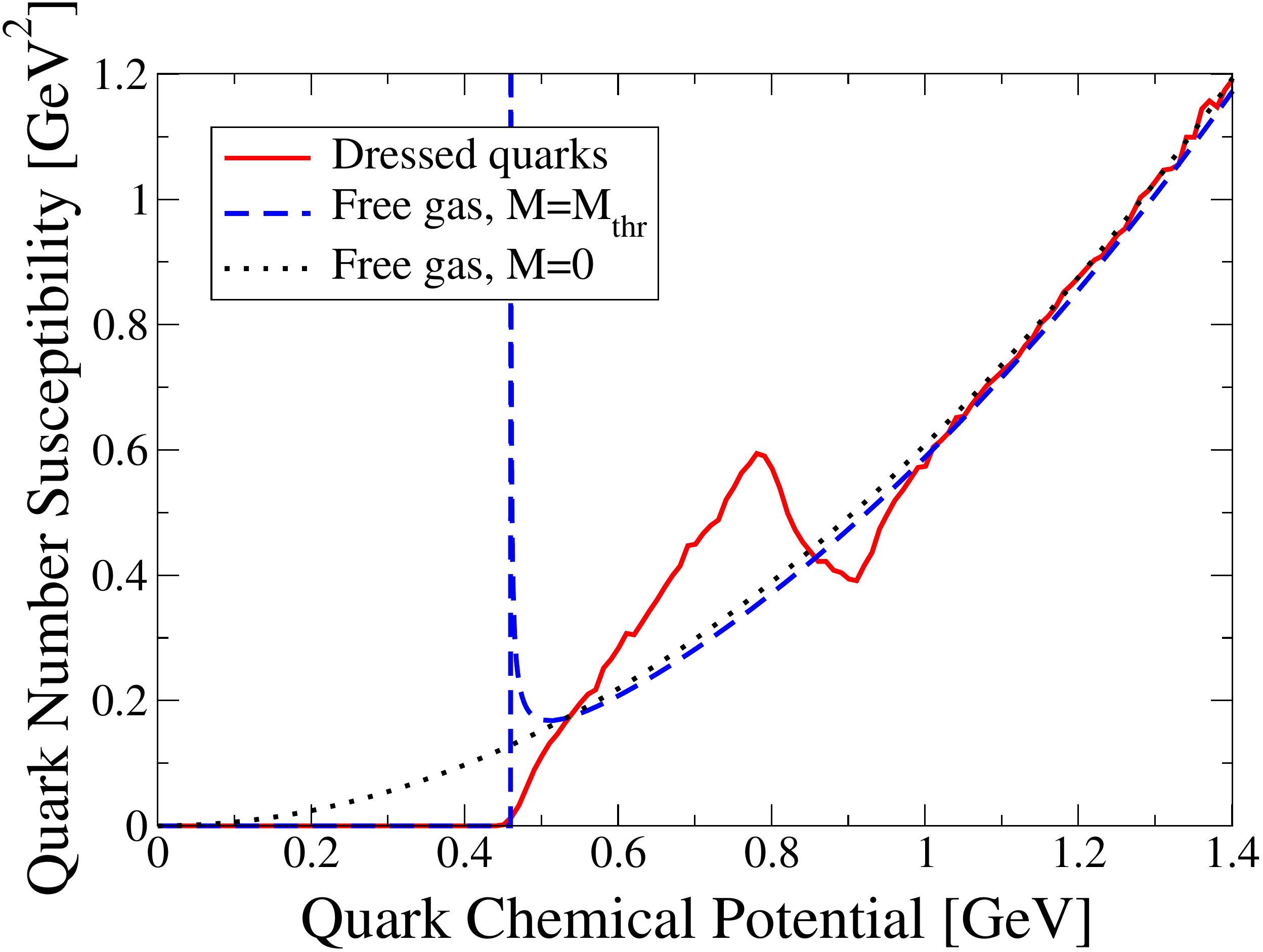}}
\caption{(a) Quark density as a function of chemical potential $\mu$ (with the results normalized by the free massless case in the detail) and (b) quark number susceptibility at $T\rightarrow0$. Notice 
the approach to the free, massless limit $\chi_{SB}=N_cN_f\mu^2/\pi^2$ 
as $\mu\rightarrow\infty$. There is a clear threshold at a scale $\mu \approx M_{thr} = 0.467$GeV.}
\label{fig4:quark_density_and_susceptibility}
\end{figure}
%%%%%%%%%%%%%%%%%%%%

%
%%%%%%%%%%%%%%%%%%%%%
%\begin{figure}
% \centering
% \subfloat[]{
% \includegraphics[width=8.5cm]{unnormalized-quark-density-zeroT.pdf}}
% \subfloat[]{\includegraphics[width=8.5cm]{unnormalized-quark-susceptibility-zeroT.pdf}}
%% \subfloat[]{\includegraphics[width=7cm]{normalized-quark-susceptibility.pdf}}
%\caption{(a) Quark density as a function of chemical potential $\mu$ and (b) quark number susceptibility at $T\rightarrow0$. Notice 
%the approach to the free, massless limit $\chi_{SB}=N_cN_f\mu^2/\pi^2$ 
%as $\mu\rightarrow\infty$. There is a clear threshold at a scale $\mu \approx M_{thr} = 0.467$GeV.}
%\label{fig4:quark_density_and_susceptibility}
%\end{figure}
%%%%%%%%%%%%%%%%%%%%%

Besides the pressure, two other relevant observables that can be calculated are the quark density (\ref{eq:quark_density}), 
and the quark number susceptibility (\ref{eq:quark_susceptibility}), as shown in Figs. (\ref{fig4:quark_density_and_susceptibility}a) 
and (\ref{fig4:quark_density_and_susceptibility}b). The behavior of these observables is  reasonably smooth and interpolates between that of free massless and massive quarks.
%In particular, one does not find any phase transition, as expected. 
The quark susceptibility is a non-monotonic function of 
the chemical potential, having an oscillation inside the window $0.8\,{\rm GeV}\lesssim\mu\lesssim1\,$GeV, resembling the result 
for the massive gas of free quarks, but with a smoothened behavior.

  Even though it is not strictly proven that a first-order phase transition occurs in cold and dense QCD, it has been predicted in several low-energy QCD models. However, no sharp, first-order transition is observed in our predictions. The massive parameters in our model, $M_3$ and $m$ are fixed by zero-temperature, zero-$\mu$ lattice data for the quark propagator. They do not depend on temperature nor chemical potential. A crossover may be reasonably described, but a first-order phase transition needs symmetry changes that, we believe, will only be attained by $T-$ and $\mu-$dependent parameters or condensates, such as $M_3(T,\mu)$. Indeed, $M_3$ can be seen as a nonlocal order parameter for chiral symmetry breaking \cite{Dudal:2013vha} and its $T,\mu$ dependence will thus play a significant role in the chiral phase transition. 

\subsection{ Results for $T\ne 0$ and $\mu\ne 0$}

In this subsection, we consider a thermal system with an imbalance between the number of fermions and anti-fermions, given by a finite $\mu$.
In Figs. (6a) and (6b), pressure and quark number density, respectively, are shown as functions of chemical potential for different values of temperature.

%%%%%%%%%%%%%%%%%%%%
\begin{figure}[!h]
\centering
 \subfloat[]{
 \includegraphics[width=7.5cm]{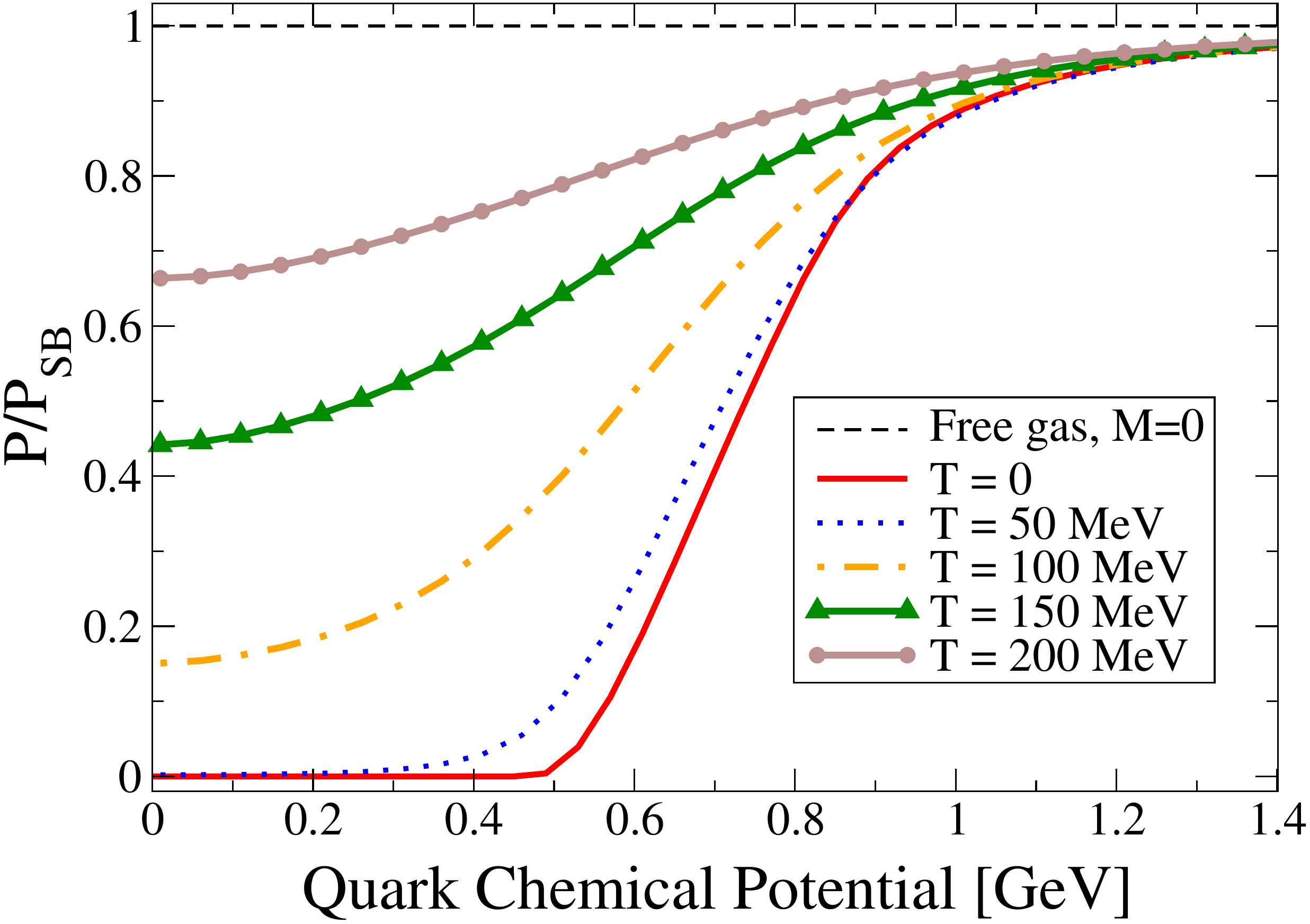}}
 \hspace{1cm}
  \subfloat[]{\includegraphics[width=7.2cm]{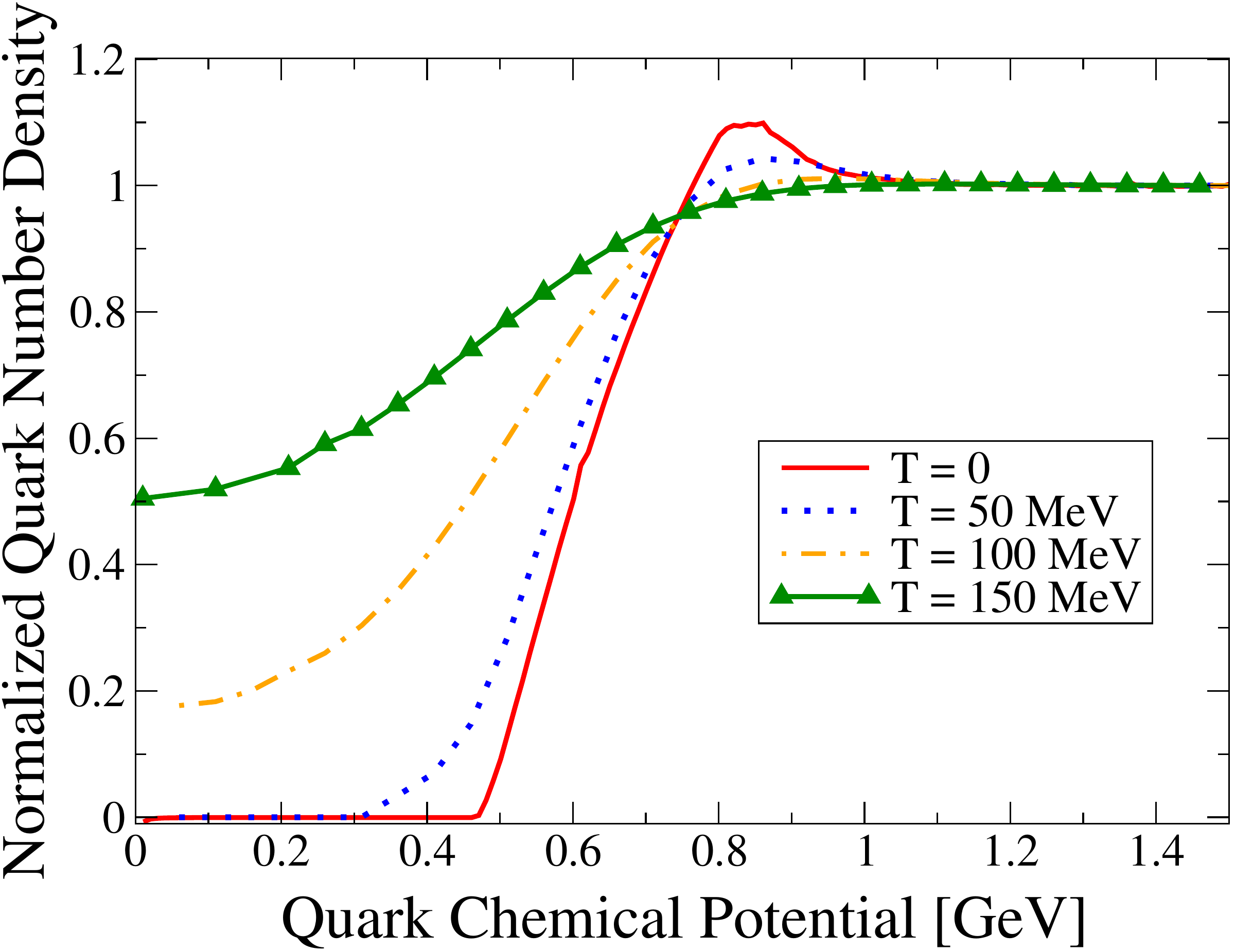}}
% \subfloat[]{\includegraphics[width=7cm]{normalized-quark-susceptibility.pdf}}
\caption{(a) Pressure and (b) quark number density (normalized by the free massless gas result) are plotted as functions of the chemical potential $\mu$ for temperature $T=\{0\textrm{ (bottom)},50,100,150,200 \textrm{ (top)}\}$ MeV.}
\label{fig6:finiteTmu}
\end{figure}
%%%%%%%%%%%%%%%%%%%%

These results further support the physical consistency of our model and the absence of thermodynamical instabilities. Moreover, we see that, as temperature increases, more thermal energy becomes available, exciting particles even at lower chemical potentials, as expected. Even though we do not see a first-order phase transition at low temperatures, as discussed above, we do observe two qualitative features generally seen in QCD models: (i) a smoothening of the transition as temperature increases and (ii) a shift of the inflection point of the curve to lower values of $\mu$.

%%%%%%%%%%%%%%%%%%%%%%%%%%%%%%%%%%%%%%%%
\section{Summary and discussion}

In this paper, we made a first exploratory study of the thermodynamics of a softly BRST broken quark model. 
A soft breaking of the BRST symmetry could in principle be present in the infrared regime of QCD without affecting any ultraviolet 
(perturbative) predictions. In this case, the BRST breaking may be intimately related 
to other nonperturbative phenomena, such as confinement and chiral symmetry breaking, representing a complementary way to 
study these issues. Evidence for this breaking has been recently found on lattice studies in the gluon sector \cite{Cucchieri:2014via}.
It also provides an argument for the absence of 
asymptotic quarks in the spectrum as well as a good analytical description of nonperturbative quark propagators as compared to lattice.

We adopt the tree-level approximation in our model, so that no explicit quark-gluon interactions are present. Nevertheless,
gluon dynamics is encoded in the nontrivial BRST-breaking background in the form of a nonlocal quark propagator.
The model reduces 
itself to a free quark model in the high energy limit, corresponding to the fermionic sector of Quantum 
Chromodynamics, and yields an infrared dynamics compatible with the available lattice data for quark propagators.
Within this model, as in other nonperturbative approaches to QCD, confinement is 
encoded in the presence of complex poles in the momentum-space quark propagator, i.e., complex masses.
The quark propagator adopted here has been explicitly shown to violate reflection positivity, being compatible with 
the absence of asymptotic quark physical states.

Despite the presence of the nontrivial BRST-breaking background, the theory is quadratic in the quark fields.
 Therefore, the  grand canonical partition function of the model could be 
calculated exactly at finite temperature and chemical potential. This allowed a computation of several 
thermodynamic quantities, as shown in the previous section. It is important to note that this model does not 
lead to thermodynamic instabilities, in spite of the presence of complex poles in propagator. We believe that the presence of three
poles (one of them real) in the propagator of our model is a crucial feature to guarantee the consistency of the thermodynamic predictions,
since quark models with pair of complex-conjugated poles alone  have displayed unphysical macroscopic behavior \cite{Benic:2012ec}.

Besides presenting a consistent thermodynamic behavior -- free of instabilities --, the model displays nontrivial
physical results, even at tree level.
At finite temperature and vanishing chemical potential, one sees a nontrivial behavior of  thermodynamic 
observables, which clearly differ from the case of free quarks at low and intermediate temperatures, 
while recovering the standard free case at high temperatures. This can be seen as an effect of the momentum-dependent
mass function, which is in turn a manifestation of the soft BRST breaking. 
It is interesting to notice that 
there is a clear rise in the pressure, as well as a peak in the trace anomaly, at around $T\sim0.15\,$GeV, 
the temperature scale of the deconfinement and chiral restoration crossovers. In spite of this, one may 
not associate this change of behavior with a phase transition, as other signatures 
are not present, as e.g. a dip in the sound velocity.

The model also allowed us to access the region of finite chemical potential at arbitrary temperatures, 
including a well-defined $T\rightarrow0$ limit.
 The finite density thermodynamics was also shown to contain nontrivial features, differing from an ideal quark gas picture.
 The system has no excitations until a threshold chemical potential 
$\mu\sim M_{thr}=0.467\,$GeV is achieved, so that the Silver-Blaze property \cite{Muroya:2003qs,Cohen:2004qp} is satisfied.
For larger chemical potentials, the limit of massless free quarks is 
approached. In spite of the clear excitation threshold $M_{thr}$, it is probably not correct to interpret the 
system as a gas of constituent quarks of mass $M_{thr}$, as can be seen, e.g., from the comparison of the 
respective pressures in Fig. (\ref{fig3:pressure_zero_temperature}). It seems more appropriate to 
resort to an interpretation in terms of an ensemble of resonances with different masses. This idea seems to be 
supported by the momentum and chemical potential dependence of the mass function (\ref{eq:mass_function_finiteT}).
As the chemical potential is increased, more phase space is made available for excitations with different momenta.
Given that the mass depends on both $\mu$ and ${\bf p}$, these newly available states have different masses as 
those present before. A similar interpretation seems to be actually possible for any $T$ or $\mu$.

Of course, further improvements and perspectives are possible and planned. 
Our results for the equation of state at large densities and low temperatures may in principle be further developed to investigate whether a 
physical compact star with a quark matter core could be created in this model.
The explicit inclusion of interactions on top of the already nontrivial nonperturbative background
may be implemented through different couplings, such as four-fermion or quark-gluon vertices, to be added perturbatively.  
Moreover, in order to explicitly compute the Polyakov loop in our model one could compute its effective potential in the background 
field formalism (for recent pure glue studies, see Refs. \cite{Canfora:2015yia,Reinosa:2014zta,Reinosa:2014ooa} and Ref.\cite{Reinosa:2015oua} for the case with heavy quarks). This way one may understand further the significant change of thermodynamic quantities observed here in terms
of the approximate order parameter for the deconfinement phase transition.

\section*{Acknowledgements}

B.W.M thanks the hospitality of the Institut f\"ur Theoretische Physik of the University of Giessen, where preliminary results of this work were presented and discussed, and FAPERJ for financial support. M.S.G. thanks the Conselho Nacional de Desenvolvimento Cient\'{\i}fico e Tecnol\'ogico (CNPq-Brazil) for financial support.  L.F.P. is supported by a BJT fellowship from the brazilian program ``Ci\^encia sem Fronteiras'' (grant number 301111/2014-6).

\begin{appendices}
  \renewcommand\thetable{\thesection\arabic{table}}
  \renewcommand\thefigure{\thesection\arabic{figure}}
  \section{Fermion effective action at finite $T$ and $\mu$} \label{app:effective-action}

%%%%%%%%%%
\subsection{Model hamiltonian}

The first step to calculate the grand partition function is to determine the hamiltonian density of the theory. It can be obtained from the lagrangian 
density in Minkowski space
\begin{equation}
  {\cal L} = {\cal L}_{Dirac}+{\cal L}_{\xi\lambda} + {\cal L}_M,
 \end{equation}
 where, omitting spinor, color, and flavor indices,
 \begin{equation}
  {\cal L}_{Dirac} = \bar\psi\,i\gamma^\mu\partial_\mu\psi - m_0\bar\psi\psi,
 \end{equation}
 and
 \begin{equation}
  {\cal L}_{\xi\lambda} = \bar\xi(-\partial^2+m^2)\lambda + \bar\lambda (-\partial^2+m^2)\xi + 
   \bar\eta(-\partial^2+m^2)\theta - \bar\theta(-\partial^2+m^2)\eta,
 \end{equation}
 and
 \begin{equation}
  {\cal L}_M = M_1^2(\bar\xi \psi + \bar\psi\xi) - M_2(\bar\lambda\psi + \bar\psi\lambda)
 \end{equation}

 The hamiltonian is the Legendre transformation of the lagrangian with respect to the generalized velocity, that is
 \begin{equation}
  {\cal H} = \sum_i \Pi_{\phi_i}(\partial_0\phi_i) - {\cal L},
 \end{equation}
 where $\Pi_{\phi_i}$ is the momentum conjugated to the field $\phi_i$. The field momenta are defined as
 \cite{Swanson:book}
 \begin{equation}\label{eq:momentum_boson}
  \Pi_{\phi_i} = \frac{\partial {\cal L}}{\partial(\partial_0\phi_i)} \;\;\;\;\mbox{for bosonic fields}
 \end{equation}
 or
 \begin{equation}\label{eq:momentum_fermion}
  \Pi_{\phi_i} = -\frac{\partial {\cal L}}{\partial(\partial_0\phi_i)} \;\;\;\;\mbox{for fermionic fields}
 \end{equation}
 After adding a total derivative to ${\cal L}_{\xi\lambda}$, one straightforwardly finds the following set of field momenta
 \begin{equation}\label{eq:field_momenta}
  \begin{array}{c|c|c|c}
    \Pi_\xi = \partial_0\bar\lambda &  \Pi_{\bar\xi} = -\partial_0\lambda & \Pi_\lambda = \partial_0\bar\xi & \Pi_{\bar\lambda} = -\partial_0\xi\\\hline
    \Pi_\theta = \partial_0\bar\eta &  \Pi_{\bar\theta} = -\partial_0\eta & \Pi_\eta = -\partial_0\bar\theta & \Pi_{\bar\eta} = \partial_0\theta
   \end{array}  
 \end{equation}
 It is important to notice that due to the first-order nature of the Dirac equation, the quark fields $\psi$ and $\bar\psi$ are 
 actually such that $i\bar\psi\gamma_0$ is the momentum conjugated to the field $\psi$, i.e.,
 \begin{equation}\label{eq:psi_momentum}
  \Pi_\psi=i\bar\psi\gamma_0%\;\;\;(\Leftrightarrow \;\;\;\bar\psi = -i\Pi_\psi\gamma_0).
 \end{equation}

 Putting all pieces together, the hamiltonian density can be written as
 \begin{equation}
  {\cal H} = {\cal H}_{Dirac} + {\cal H}_{\xi\lambda} + {\cal H}_M,
 \end{equation}
 where
 \begin{equation}
  {\cal H}_{Dirac}[\Pi_\psi,\psi] = -\Pi_\psi\gamma_0\vec\gamma\cdot\nabla\psi - im_0\Pi_\psi\gamma_0\psi,
 \end{equation}
 and
 \begin{eqnarray}
  {\cal H}_{\xi\lambda} = &&\hspace{-0.35cm}\Pi_\xi\partial_0\xi + \Pi_{\bar\xi}\partial_0\bar\xi  
  + \Pi_{\lambda}\partial_0\lambda + \Pi_{\bar\lambda}\partial_0\bar\lambda 
  + \Pi_\theta\partial_0\theta + \Pi_{\bar\theta}\partial_0\bar\theta 
  + \Pi_\eta\partial_0\eta + \Pi_{\bar\eta}\partial_0\bar\eta - {\cal L}_{\xi\lambda} \nonumber\\
  && \hspace{-.70cm} = \Pi_{\bar\xi}\Pi_{\lambda} + \Pi_{\bar\lambda}\Pi_\xi - \Pi_{\bar\theta}\Pi_{\eta} + \Pi_{\bar\eta}\Pi_{\theta} 
  - (\nabla\bar\xi)\cdot(\nabla\lambda) - (\nabla\bar\lambda)\cdot(\nabla\xi)-(\nabla\bar\eta)\cdot(\nabla\theta) + (\nabla\bar\theta)\cdot(\nabla\eta) - \nonumber\\
  &&  - m^2(\bar\xi\lambda + \bar\lambda\xi + \bar\eta\theta - \bar\theta\eta), 
 \end{eqnarray}
 and
 \begin{equation}
   {\cal H}_M = - {\cal L}_M = - M_1^2(\bar\xi\psi + \bar\psi\xi) + M_2(\bar\lambda\psi + \bar\psi\lambda).
 \end{equation}
 %

%%%
\subsection{Noether current}

In the grand canonical ensemble the average charge of the system is kept constant. This constraint can be imposed by the replacement
${\cal H}\rightarrow{\cal H} - \mu j_0$, where $j_0$ is the charge density and $\mu$ the corresponding chemical potential.
From a field theoretical point of view, such a charge density is consistently defined as the time component of the Noether current 
associated with a global $U(1)$ transformation. Let us now define one such transformation for the theory at hand 
and find the associated Noether charge.

In order to define the quark number, we first notice that the action is symmetric under the $U(1)$ transformation
 \begin{eqnarray}
  \psi\rightarrow e^{-i\alpha}\psi,\;\;\;\;\bar\psi\rightarrow e^{i\alpha}\bar\psi \nonumber\\
  \xi\rightarrow e^{-i\alpha}\xi,\;\;\;\;\bar\xi\rightarrow e^{i\alpha}\bar\xi \nonumber\\
  \lambda\rightarrow e^{-i\alpha}\lambda,\;\;\;\;\bar\lambda\rightarrow e^{i\alpha}\bar\lambda \nonumber\\
  \eta\rightarrow e^{-i\alpha}\eta,\;\;\;\;\bar\eta\rightarrow e^{i\alpha}\bar\eta \nonumber\\
  \theta\rightarrow e^{-i\alpha}\theta,\;\;\;\;\bar\theta\rightarrow e^{i\alpha}\bar\theta 
 \end{eqnarray}
 The corresponding Noether current is
 \begin{equation}
  j^\mu = \sum_i \Delta\phi_i \frac{\partial{\cal L}}{\partial(\partial_\mu\phi_i)}
 \end{equation}
 and the charge density is
 \begin{equation}
  j_0 = \sum_i \Delta\phi_i \frac{\partial{\cal L}}{\partial(\partial_0\phi_i)} = \sum_b \Delta\phi_b\,\Pi_{\phi_b} - \sum_f \Delta\phi_f\,\Pi_{\phi_f},
 \end{equation}
 where we recalled the definitions of field momenta for bosons (\ref{eq:momentum_boson}) and fermions (\ref{eq:momentum_fermion}).
 The explicit expression for the charge density in terms of fields and field momenta is
 \begin{equation}
  j_0 = i\psi\Pi_\psi +i\xi\Pi_\xi - i\bar\xi\Pi_{\bar\xi} + i\lambda\Pi_\lambda - i\bar\lambda\Pi_{\bar\lambda} 
  - i\theta\Pi_\theta + i\bar\theta\Pi_{\bar\theta} - i\eta\Pi_\eta + i\bar\eta\Pi_{\bar\eta}.
 \end{equation}

%%%%
\subsection{Quark in-medium effective action}

 In the functional integral formalism, the partition function is written as
 \begin{eqnarray}\label{eq:partition_function_path_integral}
  Z(T,\mu) = &&\hspace{-0.44cm}{\rm Tr}\exp\left(-\beta\hat H + \beta\mu\hat Q\right)\nonumber\\
  =&&\hspace{-0.44cm}\int[D\Pi_\psi][D\psi][D\xi][D\Pi_\xi][D\bar\xi][D\Pi_{\bar\xi}][D\lambda][D\Pi_\lambda][D\bar\lambda][D\Pi_{\bar\lambda}]
  [D\eta][D\Pi_\eta][D\bar\eta][D\Pi_{\bar\eta}][D\theta][D\Pi_\theta][D\bar\theta][D\Pi_{\bar\theta}]\nonumber\\
  &&\times \exp\left[\int_0^\beta d^4x\left(\sum_i i\Pi_i(\partial_4\phi_i) - {\cal H} + \mu j_4\right)\right],
 \end{eqnarray}
 where now the fields have been analytically continued to imaginary time according to $x_0\rightarrow x_4=ix_0$. The constraint of finite 
 temperature is imposed on the system by taking the time direction compact, i.e., $0\leq x_4\leq\beta\equiv1/T$. This is expressed in the integration
 symbol
 \begin{equation}
  \int_0^\beta\,d^4x\,(\cdots) \equiv \int_0^\beta\,dx_4\int\,d^3x\,(\cdots).
 \end{equation}

 Notice that one must start from 
 the expression (\ref{eq:partition_function_path_integral}), where integration over field momenta must still be made. As is well known from the study 
 of a complex scalar field (see, e.g., \cite{KapustaGale}), the field momentum integration in the presence of a chemical potential does not simply lead 
 to the lagrangian plus a $\mu j_4$ term. Rather, the result of including a chemical potential is typically to shift the time derivative as 
 $\partial_4\rightarrow(\partial_4-\mu)$.
 
 The integrand of the exponent of (\ref{eq:partition_function_path_integral}) is 
 \begin{eqnarray}
  I:=&&\hspace{-0.33cm}\sum_k i\Pi_k\dot\phi_k - {\cal H} + \mu j_4 \nonumber\\
     =&&\hspace{-0.33cm} i\Pi_\psi\dot\psi + i\Pi_\xi\dot\xi + i\Pi_{\bar\xi}\dot{\bar\xi} + i\Pi_\lambda\dot\lambda + i\Pi_{\bar\lambda}\dot{\bar\lambda} 
     + i\Pi_\theta\dot\theta + i\Pi_{\bar\theta}\dot{\bar\theta} + i\Pi_\eta\dot\eta + i\Pi_{\bar\eta}\dot{\bar\eta}+\nonumber\\
     &&\hspace{-0.33cm} + \Pi_\psi\gamma_0\vec\gamma\cdot\nabla\psi + im_0\Pi_\psi\gamma_0\psi - \Pi_{\bar\xi}\Pi_{\lambda} - \Pi_{\bar\lambda}\Pi_\xi + \Pi_{\bar\theta}\Pi_{\eta} - \Pi_{\bar\eta}\Pi_{\theta}  + \nonumber\\
     &&\hspace{-0.33cm} + (\nabla\bar\xi)\cdot(\nabla\lambda) + (\nabla\bar\lambda)\cdot(\nabla\xi) +(\nabla\bar\eta)\cdot(\nabla\theta) - (\nabla\bar\theta)\cdot(\nabla\eta) + \nonumber\\
     &&\hspace{-0.33cm} + m^2(\bar\xi\lambda + \bar\lambda\xi + \bar\eta\theta - \bar\theta\eta) + M_1^2(\bar\xi \psi + \bar\psi\xi) - M_2(\bar\lambda\psi + \bar\psi\lambda) +\nonumber\\
     &&\hspace{-0.33cm} - i\mu\Pi_\psi\psi - i\mu\Pi_\xi\xi + i\mu\Pi_{\bar\xi}\bar\xi - i\mu\Pi_\lambda\lambda + i\mu\Pi_{\bar\lambda}\bar\lambda - \nonumber\\
     && - i\mu\theta\Pi_\theta + i\mu\bar\theta\Pi_{\bar\theta} - i\mu\eta\Pi_\eta + i\mu\bar\eta\Pi_{\bar\eta} \nonumber\\
     =&&\hspace{-0.33cm}i\Pi_\psi\left(\dot\psi - \mu\psi -i\gamma_0\vec\gamma\cdot\nabla\psi + m_0\gamma_0\psi\right) + i\Pi_{\bar\xi}(\dot{\bar\xi}+\mu\bar\xi)  + i\Pi_\lambda(\dot\lambda - \mu\lambda) + \nonumber\\
     &&\hspace{-0.44cm} + i\Pi_{\bar\lambda}(\dot{\bar\lambda}+\mu\bar\lambda) + i\Pi_\xi(\dot\xi - \mu\xi) + i\Pi_{\bar\theta}(\dot{\bar\theta}+\mu\bar\theta) + i\Pi_\eta(\dot\eta - \mu\eta) + \nonumber\\
     &&\hspace{-0.44cm} + i\Pi_{\bar\eta}(\dot{\bar\eta} +\mu\bar\eta) + i\Pi_\theta(\dot\theta - \mu\theta) - \Pi_{\bar\xi}\Pi_{\lambda} - \Pi_{\bar\lambda}\Pi_\xi + \Pi_{\bar\theta}\Pi_{\eta} - \Pi_{\bar\eta}\Pi_{\theta} + \nonumber\\
     &&\hspace{-0.44cm} + (\nabla\bar\xi)\cdot(\nabla\lambda) + (\nabla\bar\lambda)\cdot(\nabla\xi) + (\nabla\bar\eta)\cdot(\nabla\theta) - (\nabla\bar\theta)\cdot(\nabla\eta) +  \nonumber\\
     &&\hspace{-0.44cm} + m^2(\bar\xi\lambda + \bar\lambda\xi + \bar\eta\theta - \bar\theta\eta) + M_1^2(\bar\xi \psi + \bar\psi\xi) - M_2(\bar\lambda\psi + \bar\psi\lambda)
 \end{eqnarray}
 where we used the notation $(\dot\phi_k \equiv \partial_4\phi_k)$.
 
 Note that we can separetely integrate the bosonic and the fermionic auxiliary fields. As a result, we shall find an effective action for the fermion fields. Interestingly
 enough, we find a similar result as compared to the vacuum nonlocal action, Eq. (\ref{eq:non-local-action}), the only difference being the 
 replacement\footnote{Notice that this is the same as in the case of Dirac fermions or commuting complex scalar fields in the presence of a chemical potential.} 
 $\partial_4^2\rightarrow -(i\omega_n+\mu)^2$. We now demonstrate this result, while we leave the integral in the quark field $\psi$ to the main text. 

  Let us first consider the {\it anticommuting} auxiliary fields $(\xi,\bar\xi,\lambda,\bar\lambda)$. The relevant part of the integrand for the field momentum integral is
 \begin{eqnarray}
  I_{\Pi_f}:=&&\hspace{-0.35cm}  - \Pi_{\bar\xi}\Pi_{\lambda} + i\Pi_{\bar\xi}(\dot{\bar\xi}+\mu\bar\xi)  + i\Pi_\lambda(\dot\lambda - \mu\lambda)  
     - \Pi_{\bar\lambda}\Pi_\xi  + i\Pi_{\bar\lambda}(\dot{\bar\lambda}+\mu\bar\lambda) + i\Pi_\xi(\dot\xi - \mu\xi) \nonumber\\     
     =&&\hspace{-0.44cm} -\left[ \Pi_{\bar\xi} + i\left(\dot\lambda - \mu\lambda\right)\right]\left[ \Pi_{\lambda} - i\left(\dot{\bar\xi} + \mu\bar\xi\right)\right] - 
       \left[ \Pi_{\bar\lambda} + i\left(\dot\xi - \mu\xi\right)\right]\left[ \Pi_{\xi} - i\left(\dot{\bar\lambda} + \mu\bar\lambda\right)\right] - \nonumber\\
      &&- \left(\dot{\bar\xi} + \mu\bar\xi\right)\left(\dot\lambda - \mu\lambda\right) - \left(\dot{\bar\lambda} + \mu\bar\lambda\right)\left(\dot\xi - \mu\xi\right)
 \end{eqnarray}
 Notice that the field variables and momenta above are Grassmann variables. Once this expression is under a functional integral, we may shift the momenta in $I_f$ 
 without changing the integral, so that 
 \begin{eqnarray}\label{eq:momentum_integral_aux_fermions}
  Z_{\Pi_f}:=\int[D\Pi_\xi][D\Pi_{\bar\xi}][D\Pi_{\lambda}][D\Pi_{\bar\lambda}]e^{\int_0^\beta\,d^4x\,I_f} && \nonumber\\
  && \hspace{-6cm}= e^{-\int_0^\beta\,d^4x\, \left[\left(\dot{\bar\xi} + \mu\bar\xi\right)\left(\dot\lambda - \mu\lambda\right) + \left(\dot{\bar\lambda} + \mu\bar\lambda\right)\left(\dot\xi - \mu\xi\right)\right]}
  \int[D\Pi_\xi][D\Pi_{\bar\xi}][D\Pi_{\lambda}][D\Pi_{\bar\lambda}]e^{-\int_0^\beta\,d^4x\,\left(\Pi_{\bar\xi}\Pi_\lambda + \Pi_{\bar\lambda}\Pi_\xi\right)} \nonumber\\
  &&\hspace{-6cm}= e^{-\int_0^\beta\,d^4x\, \left[\left(\dot{\bar\xi} + \mu\bar\xi\right)\left(\dot\lambda - \mu\lambda\right) + \left(\dot{\bar\lambda} + \mu\bar\lambda\right)\left(\dot\xi - \mu\xi\right)\right]}
 \end{eqnarray}
 Let us now compute the momentum integral of the bosonic auxiliary fields. The relevant integrand is
 \begin{eqnarray}
  \hspace{-0.55cm}I_{\Pi_b} =&&\hspace{-0.35cm} \Pi_{\bar\theta}\Pi_\eta + i\Pi_{\bar\theta}\left(\dot{\bar\theta} + \mu\bar\theta\right) + i\Pi_{\eta}\left(\dot{\eta} - \mu \eta\right) 
       - \Pi_{\bar\eta}\Pi_\theta + i\Pi_{\bar\eta}\left(\dot{\bar\eta} + \mu\bar\eta\right) + i\Pi_{\theta}\left(\dot{\theta} - \mu \theta\right)\nonumber\\
      =&&\hspace{-0.35cm}\left[\Pi_{\bar\theta} + i(\dot\eta - \mu\eta)\right]\left[\Pi_\eta + i \left(\dot{\bar\theta} + \mu\bar\theta\right)\right] - \left[\Pi_{\bar\eta} - i(\dot\theta - \mu\theta)\right]\left[\Pi_\theta - i \left(\dot{\bar\eta} + \mu\bar\eta\right)\right] + \nonumber\\
       &&  + \left(\dot{\bar\theta} + \mu\bar\theta\right)\left(\dot\eta - \mu\eta\right) - \left(\dot{\bar\eta} + \mu\bar\eta\right)\left(\dot\theta - \mu\theta\right)
 \end{eqnarray}
 As in the fermionic case, for fixed field configurations, one is allowed to shift the field momenta by any arbitrary function without changing the functional integral. As a result,
 \begin{eqnarray}\label{eq:momentum_integral_aux_bosons}
  Z_{\Pi_b}:=&&\hspace{-0.35cm}\int[D\Pi_\eta][D\Pi_{\bar\eta}][D\Pi_\theta][D\Pi_{\bar\theta}]e^{\int_0^\beta\,d^4x\,I_b}\nonumber\\
  =&& \hspace{-0.35cm} e^{\int_0^\beta\,d^4x\,\left[\left(\dot{\bar\theta} + \mu\bar\theta\right)\left(\dot\eta - \mu\eta\right) + \left(\dot{\bar\eta} + \mu\bar\eta\right)\left(\dot\theta - \mu\theta\right)\right]}
  \int[D\Pi_\eta][D\Pi_{\bar\eta}][D\Pi_\theta][D\Pi_{\bar\theta}]e^{\int_0^\beta\,d^4x\,\left[\Pi_{\bar\theta}\Pi_\eta + \Pi_{\bar\eta}\Pi_\theta\right]} \nonumber\\
  =&&\hspace{-0.35cm} e^{\int_0^\beta\,d^4x\,\left[\left(\dot{\bar\theta} + \mu\bar\theta\right)\left(\dot\eta - \mu\eta\right) - \left(\dot{\bar\eta} + \mu\bar\eta\right)\left(\dot\theta - \mu\theta\right)\right]}
 \end{eqnarray}
 In our next step, we integrate over the auxiliary boson fields $(\theta,\bar\theta, \eta,\bar\eta)$. The relevant integrand is
 \begin{eqnarray}
  I_b:=&&\hspace{-0.35cm}\left(\dot{\bar\theta} + \mu\bar\theta\right)\left(\dot\eta - \mu\eta\right) - \left(\dot{\bar\eta} + \mu\bar\eta\right)\left(\dot\theta - \mu\theta\right) + 
   (\nabla\bar\eta)\cdot(\nabla\theta) - (\nabla\bar\theta)\cdot(\nabla\eta) + m^2(\bar\eta\theta - \bar\theta\eta) \nonumber\\
  =&&\hspace{-0.35cm} \dot{\bar\theta}\dot\eta - \mu\dot{\bar\theta}\eta + \mu\bar\theta\dot\eta - \mu^2\bar\theta\eta + (\nabla\bar\eta)\cdot(\nabla\theta) + m^2\bar\eta\theta - 
   - \left[\dot{\bar\eta}\dot\theta - \mu\dot{\bar\eta}\theta + \mu\bar\eta\dot\theta - \mu^2\bar\theta\eta + (\nabla\bar\theta)\cdot(\nabla\eta) + m^2\bar\theta\eta\right] \nonumber\\
  \longrightarrow&&\hspace{-0.35cm} \bar\theta\left(-\partial_0^2 + 2\mu\partial_0 - \mu^2 - \nabla^2 + m^2\right)\eta - \bar\eta\left(-\partial_0^2 + 2\mu\partial_0 - \mu^2 - \nabla^2 + m^2\right)\theta
 \end{eqnarray}
 where, in the last step we added a total derivative to the integrand.
 The resulting functional integral is (aside from an infinite numerical constant)
 \begin{eqnarray}
  \int[D\bar\theta][D\theta][D\bar\eta][D\eta]e^{\int_0^\beta\,d^4x\,I_b} &=& \det[-D^2 + m^2]^{-\frac12(4N_c)} \det[D^2 - m^2]^{-\frac12(4N_c)} \nonumber\\
  &=& \det[-D^2+m^2]^{-4N_c},
 \end{eqnarray}
 where we defined the operator $D^2:=(\partial_4 - \mu)^2 + \nabla^2$, which corresponds to 
 $(i\omega_n + \mu)^2 - {\bf p}^2$ in momentum space.
 At this point, it should be clear that the final result should be the same as the one at zero temperature (in euclidean space), with the replacement
 $p_4\rightarrow \omega_n-i\mu$. Let us continue our calculation in order to explicitly check that this is the case.
 The relevant integrand in the fermionic auxiliary fields is
 \begin{eqnarray}
  I_f:=&&\hspace{-0.35cm} \left(\dot{\bar\xi} + \mu\bar\xi\right)\left(\dot\lambda - \mu\lambda\right) + \left(\dot{\bar\lambda} + \mu\bar\lambda\right)\left(\dot\xi - \mu\xi\right) 
   + (\nabla\bar\xi)\cdot(\nabla\lambda) + (\nabla\bar\lambda)\cdot(\nabla\xi) + m^2(\bar\xi\lambda + \bar\lambda\xi) +   \nonumber\\
  &&  + M_1^2(\bar\xi \psi + \bar\psi\xi) - M_2(\bar\lambda\psi + \bar\psi\lambda) \nonumber\\
  &&\hspace{-1cm}\rightarrow \; \bar\xi\left(-D^2+m^2\right)\lambda + \bar\lambda\left(-D^2+m^2\right)\xi + M_1^2(\bar\xi \psi + \bar\psi\xi) - M_2(\bar\lambda\psi + \bar\psi\lambda),
 \end{eqnarray}
 where we added a total derivative and used the definition of $D^2$ in the last step.

 The resulting functional integral is again not altered if we add constant functions to the integrated fields. Once the $\psi$ field is kept fixed in the 
 integration of $(\bar\xi,\xi,\bar\lambda,\lambda)$, we may freely make the shifts
 \begin{eqnarray}
  &&\xi\;\longrightarrow\;\xi + M_2\Delta^{-1}\psi\;\;\;\;\;\;\;\;\;\;\;\bar\xi\;\longrightarrow\;\bar\xi + M_2\bar\psi\Delta^{-1} \nonumber\\
  &&\lambda\;\longrightarrow\;\lambda - M_1^2\Delta^{-1}\psi\;\;\;\;\;\;\;\;\;\;\;\bar\lambda\;\longrightarrow\;\bar\lambda - M_1^2\bar\psi\Delta^{-1} \nonumber\\
 \end{eqnarray}
 where $\Delta:=-D^2+m^2$.
 It follows that
 \begin{eqnarray}
  I_f=&&\hspace{-0.35cm} \left(\bar\xi + M_2\bar\psi\Delta^{-1}\right)\Delta \left(\lambda - M_1^2\Delta^{-1}\psi\right) 
      + \left(\bar\lambda - M_1^2\bar\psi\Delta^{-1}\right)\Delta\left(\xi + M_2\Delta^{-1}\psi\right) + \nonumber\\
      &&\hspace{-0.33cm} + M_1^2\left(\bar\xi + M_2\bar\psi\Delta^{-1}\right)\psi + M_1^2\bar\psi\left(\xi + M_2\Delta^{-1}\psi\right) 
      - M_2\left(\bar\lambda - M_1^2\bar\psi\Delta^{-1}\right)\psi - M_2\bar\psi\left(\lambda - M_1^2\Delta^{-1}\psi\right)\nonumber\\
      &&\hspace{-0.55cm} = \bar\xi \Delta \lambda + \bar\lambda \Delta \xi + 2M_1^2M_2\bar\psi \Delta^{-1}\psi 
 \end{eqnarray}
 The integral in the fermionic auxiliary fields is then
 \begin{eqnarray}
  \int[D\bar\xi][D\xi][D\bar\lambda][D\lambda]e^{-\int_0^\beta\,d^4x\,I_f} &=& \det[\Delta]^{4N_c}e^{-\int_0^\beta\,d^4x\,2M_1^2M_2\bar\psi \left(\Delta^{-1}\right)\psi}\nonumber\\
   &=& \det[-D^2+m^2]^{4N_c}e^{-\int_0^\beta\,d^4x\,2M_1^2M_2\bar\psi \left(\frac{1}{-D^2+m^2}\right)\psi}
 \end{eqnarray}
 We now finally arrive at our expression for the partition function in terms of the quark fields only
 \begin{equation}
  Z(T,\mu) = \int[D\Pi_\psi][D\psi]\exp\left[\int_0^\beta\,d^4x\,i\Pi_\psi\gamma_4\left(\gamma_4(\partial_4 - \mu) -i\vec\gamma\cdot\nabla + \hat M_p\right)\psi \right]
 \end{equation}
 where the nonlocal mass operator is
 \begin{equation}
  \hat M_p = \frac{2M_1^2M_2}{-D^2+m^2} + m_0 \rightarrow \frac{2M_1^2M_2}{-(i\omega_n+\mu)^2 + {\bf p}^2 + m^2} + m_0
 \end{equation}
 Finally, after the change of variables\footnote{This changes the integration measure by a constant, which we absorb in the definition of $Z$.} 
 $\Pi_\psi\rightarrow \bar\psi=-i\Pi_\psi\gamma_4$, the partition function can be written as 
 \begin{equation}
  Z(T,\mu) = \int[D\bar\psi][D\psi]\exp\left[-\int_0^\beta\,d^4x\,{\cal L}_{nl}[\bar\psi,\psi]\right] 
 \end{equation}
 where
 \begin{equation}
  {\cal L}_{nl}[\bar\psi,\psi]=\bar\psi\left[\gamma_4(\partial_4-\mu) - i\vec\gamma\cdot\nabla + \frac{2M_1^2M_2}{-D^2+m^2} + m_0\right]\psi
 \end{equation}
 is the (nonlocal) quark effective lagrangian at finite temperature and chemical potential.

\end{appendices}

%%%%%%%%

\end{document}